\documentclass[manuscript]{aastex}

\slugcomment{To be submitted to ApJ}

\shorttitle{Modeling of M-dwarf GJ832}
\shortauthors{Fontenla et al.}

\begin{document}


\title{SEMI-EMPIRICAL MODELING OF THE PHOTOSPHERE, CHROMOSPHERE, 
TRANSITION REGION, AND CORONA OF THE 
M-DWARF HOST STAR GJ 832\footnote{Based on observations made with the NASA/ESA
Hubble Space Telescope obtained from the Data Archive at the Space
Telescope Science Institute, which is operated by the Association of
Universities for Research in Astronomy, Inc., under NASA contract NAS
AR-09525.01A. These observations are associated with programs \#12034,
12035, 12464.}}


\author{J.M. Fontenla\altaffilmark{2}, Jeffrey  L. Linsky\altaffilmark{3}, 
Jesse Witbrod\altaffilmark{4}, Kevin France\altaffilmark{5},
 A. Buccino\altaffilmark{6,7}, Pablo Mauas\altaffilmark{6},
 Mariela Vieytes\altaffilmark{6}, and Lucianne M. Walkowicz\altaffilmark{8}}

\altaffiltext{2}{NorthWest Research Associates, Boulder, CO 80301;
  {\tt johnf@digidyna.com}}

\altaffiltext{4}{University of Colorado Boulder, CO 80301;
{\tt jesse.witbrod@colorado.edu}}

\altaffiltext{3}{JILA, University of Colorado and NIST, Boulder, CO 
80309-0440; {\tt jlinsky@jila.colorado.edu}}

\altaffiltext{4}{University of Colorado Boulder, CO 80301;
{\tt jesse.witbrod@colorado.edu}}

\altaffiltext{5}{LASP, University of Colorado Boulder, CO 80309-0600;
{\tt kevin.france@lasp.colorado.edu}}

\altaffiltext{6}{Instituto de Astronom\'ia y F\'isica del Espacio 
(CONICET-UBA), C.C. 67, Sucursal 28, C1428EHA, Buenos Aires, Argentina; 
{\tt abuccino@iafe.uba.ar, pablo@iafe.uba.ar, mariela@iafe.uba.ar}}

\altaffiltext{7}{Visiting Astronomer,
    Complejo Astron\'omico El Leoncito operated under agreement
    between the Consejo Nacional de Investigaciones Cient\'\i ficas y
    T\'ecnicas de la Rep\'ublica Argentina and the National
    Universities of La Plata, C\'ordoba and San Juan.}

\altaffiltext{8}{The Adler Planetarium, Chicago, IL 60605; 
{\tt LWalkowicz@adlerplanetarium.org}}



\begin{abstract}

Stellar radiation from X-rays to the visible provides the energy
that controls the photochemistry and mass loss from exoplanet 
atmospheres. The important
extreme ultraviolet (EUV) region (10--91.2~nm) is inaccessible and should be 
computed from a reliable stellar model. It is
essential to understand the formation regions and physical processes
responsible for the various stellar emission features in order to
predict how the spectral energy distribution varies with age and
activity levels. We compute a state-of-the-art 
semi-empirical atmospheric model and
the emergent high-resolution synthetic spectrum of the 
moderately active M2~V star GJ~832
as the first of a series of models for stars with different activity
levels. Using non-LTE radiative transfer techniques and including many
molecular lines, we construct a one-dimensional simple model for 
the physical structure of the star's chromosphere, chromosphere-corona
transition region, and corona. The synthesized 
spectrum for this model fits the continuum and
lines across the UV to optical spectrum. 
Particular emphasis is given to the 
emission lines at wavelengths shorter than 300~nm observed with
{\em HST}, which have important effects on 
the photochemistry in the exoplanet atmospheres. 
The FUV line ratios indicate that the transition region of GJ~832 
is more biased to hotter material than that of the quiet Sun. 
The excellent agreement of our computed EUV luminosity with that
obtained by two other techniques indicates that our model predicts 
reliable EUV emission from GJ~832.
We find that unobserved EUV flux of GJ~832, which heats the outer
atmospheres of exoplanets and drives their mass loss, is
comparable to the active Sun.

\end{abstract}





\section{INTRODUCTION AND MOTIVATION}

The discovery of large numbers of exoplanets by the {\em Kepler} 
\citep{Koch10} and
{\em CoRoT} \citep{Baglin03} satellites and by ground-based 
radial velocity surveys 
has stimulated enormous interest in determining the properties and 
radiative output of M dwarf host stars. The low mass and luminosity of 
M dwarfs makes these stars favorable targets for the discovery and 
characterization of exoplanets in their habitable zones 
\citep[e.g.,][]{Kasting93,Scalo07,Tarter07,Kopparapu13}. 
These exoplanets are potentially easier 
to discover because they have large orbital radial velocities and relatively
high probabilities for transits as a consequence of their being 
much closer to their host stars than for solar-type stars. 
Because M dwarf stars are much smaller and cooler than solar-type stars,
their exoplanets occult a larger fraction of the stellar surface 
during transits with higher contrast between the exoplanet's faint
absorption and emission spectra and the star's emission.
Unfortunately, variability of the stellar
UV and optical flux due to rotation and starspots complicates the 
analysis of transit light curves for exoplanets of M dwarf stars 
\citep[e.g.,][]{Llama15} and even the detection of habitable-zone exoplanets
\citep{Newton16a}.

\citet{Dressing15} extracted from the {\em Kepler} data the 
mean number of 
Earth-size (0.5--1.4 $R_{E}$) exoplanets with orbital periods shorter 
than 50 days, corresponding to orbital distances that place many 
exoplanets inside the habitable zones of M dwarfs. They found that 
the occurrence rate of such planets is $0.16^{+0.17}_{-0.07}$ per 
cool dwarf and predicted with 95\% confidence that the 
nearest nontransiting Earth-sized planet in its habitable zone 
is located within 5~pc of the Sun. \citet{Kopparapu13} 
proposed that the occurrence rate for such planets is about a factor of three 
larger based on new calculations of the habitable zone boundaries, 
which will place the likely location of the nearest habitable planet 
even closer than 5~pc. These predictions are powerful arguments for 
studying M-dwarf stars, the closest and most numerous stars in the Galaxy, 
whether or not they are presently known to harbor exoplanets. The upcoming 
{\em Transiting Exoplanet Survey Satellite (TESS)} and the 
{\em James Webb Space Telescope (JWST)} missions will play major roles 
in these studies. 

The many observational studies of the optical H$\alpha$ and Ca~II H and
K lines \citep[e.g.,][]{Stauffer86,Robinson90,Houdebine97,Mauas00,
Walkowicz09,Perez14} 
have until now provided the principle diagnostics for modeling the magnetically 
heated gas in the chromospheres of low-mass stars. The presence and
strength of H$\alpha$ emission is the commonly used indicator of 
activity levels in M stars 
and even brown dwarfs \citep{Berger10}. While the cores of
the Ca~II lines increase monotonically with magnetic heating and provide
an unambiguous measure of stellar activity, the H$\alpha$ line behaves
differently with increasing activity. 
\citet{Cram87}, \citet{Houdebine97}, and later modelers have predicted
that with increasing chromospheric heating the H$\alpha$
absorption line, which is very weak in the least active cool stars, 
first becomes a deeper absorption line and then 
fills-in to become an emission line. An
accurate measure of the correlation of the fluxes or equivalent widths
of the H$\alpha$ and Ca~II lines requires simultaneous
observations of the Ca~II and H$\alpha$ lines, especially for the 
highly variable M dwarfs. \citet{Walkowicz09} obtained 
simultaneous high-resolution spectra of the Ca~II and H$\alpha$ lines 
in M3~V stars. They observed a strong positive correlation 
of H$\alpha$ and Ca~II emission in active stars but only a weak or nonexistent
correlation for inactive and moderately active stars, because for a
large range of Ca~II equivalent widths there is only a very small
range of in H$\alpha$ equivalent widths.
This uncertain correlation for low activity stars, which has been noted 
by previous observers, 
\citep[e.g.,][]{Robinson90,Buccino11,Buccino14}), highlights the need for
spectral diagnostics of emission features that all have a monotonic 
dependence on activity and thus should be positively correlated. 
An example is the correlation of the
chromospheric Mg~II emission lines (similar to the
Ca~II emission lines) with X-ray emission for M stars as found by
\citet{Stelzer13} and others.

X-ray observations of M-dwarf coronae show a nearly four orders-of- 
magnitude range in X-ray luminosity (L$_X$) between the least and 
most active M dwarfs and a correlation of L$_X$ with X-ray spectral 
hardness and thus with higher coronal temperatures 
\citep{Schmitt95,Fleming95}. 
\citet{Sanz-Forcada11} has computed the total X-ray and 
extreme-ultraviolet flux for many stars, including GJ~832, from their 
observed X-ray spectra and developed scaling laws for this flux as a
function of stellar age. During flares, M dwarfs show both 
enhanced X-ray emission and hotter coronal plasma \citep{Robrade05}.
Searches for effects of exoplanets on the chromospheres and 
coronae of their host stars have provided some evidence for
stellar-planetary interactions 
\citep{Shkolnik03,Lanza08,Shkolnik05,Kashyap08,
Poppenhaeger10,Poppenhaeger14}.
Very recently, \citet{France16} found that additional heating in host
star's chromosphere-corona transition regions (hereafter called
transition regions) is correlated with the presence of a 
close-in exoplanet.

High-resolution ultraviolet (UV) spectroscopy provides a wide range 
of diagnostics for computing semiempirical models of M dwarfs 
from the top of the photosphere through the chromosphere and 
into the corona. An excellent example of such a spectrum is the 
spectral atlas of the young dM1e star AU Mic obtained with the 
E140M grating (resolution $\lambda/\Delta\lambda = 45, 000$) of the 
Space Telescope Imaging Spectrograph (STIS) on the {\em Hubble 
Space Telescope (HST)}. In this spectrum covering the wavelength 
region 117.5--170~nm, \citet{Pagano00} identified 142 emission 
lines of 28 species extending from chromospheric neutrals 
(e.g., C~I and O~I) and ions (Fe~II and Si~II), through ions 
formed in the transition region between 20,000 K and 200,000 K 
(C II--IV, Si III--IV, O III--V, and N~V), and into the corona 
(Fe~XXI 135.4~nm line formed at $10^7$~K). With this rich group 
of emission lines, \citet{Pagano00} were able to construct 
an emission-measure distribution for the stellar chromosphere 
and transition region and to compute the electron density 
in the transition region from O IV intersystem line ratios. 

The MUSCLES (Measurements of the Uv Spectral Characteristics of Low Mass
Exoplanetary Systems) observing program \citep{France13} used both 
STIS and the Cosmic Origins Spectrograph (COS) on {\em HST} 
to obtain 115--314~nm 
spectra of six weakly active M dwarfs hosting exoplanets, including 
GJ~832. These spectra have resolutions of 7.5--17 km~s$^{-1}$ in the 
115--179.5~nm region and 2.6 km~s$^{-1}$ in the near UV 
(NUV, 170--300~nm) including the Mg~II h and k lines near 280~nm. 
An important result of this 
program is that while the NUV spectra of M dwarfs are a factor 
of $10^3$ times fainter than the Sun at the habitable zone distance 
from the star, the far-UV (FUV, 115--170~nm) spectra formed in the 
chromospheres of M dwarfs have fluxes comparable to or brighter than
the Sun \citep{France12}.
The reconstructed Lyman-alpha (Ly-$\alpha$) emission line 
corrected for interstellar absorption in M dwarf spectra 
has flux comparable to the entire rest of the 116--305~nm spectrum
\citep{France13}.
Fluxes for these stars, including the reconstructed Ly-$\alpha$ 
flux, are included in the MAST
website.\footnote{https://archive.stsci.edu/prepds/muscles/} 

The MUSCLES Treasury Survey extends the previous MUSCLES pilot 
observing program to include quasi-simultaneous 
X-ray through near-IR observations of 7 M and 4 K dwarf host stars located 
within 20~pc of the Sun.
\citet{France16} present an overview of the survey showing
examples of the spectra and spectral energy distributions (SEDs) between
0.5~nm and 5.5$\mu$m. \citet{Youngblood16} compute reconstructions of
the Lyman-$\alpha$ lines and use the reconstructed Lyman-$\alpha$ fluxes to
estimate the unobservable extreme ultraviolet fluxes (EUV, 10-91.2~nm) of
the 11 stars. In the third of the initial survey papers,
\citet{Loyd16} describe the stitching together of the different data
sets that constitute the spectral energy distributions (SEDs) and compute
photodissociation spectra of different molecules given
the unattenuated SEDs of three host stars.

The observed UV spectra of M dwarfs are essential input 
for the computation of photochemical models of exoplanet atmospheres. 
The NUV stellar flux dissociates O$_2$, O$_3$, and other
molecules, while the FUV flux dissociates H$_2$O, CO$_2$, CH$_4$, 
and other molecules. The stellar particle flux during a large flare 
can even deplete much of a planetary atmosphere's ozone 
\citep{Segura10}. The very bright Ly-$\alpha$ line plays a major 
role in the photochemical processes. For example, photochemical models 
for the mini-Neptune GJ~436b \citep{Miguel15}, using the MUSCLES 
data as input, show that the flux of Ly-$\alpha$ and other lines 
in the stellar FUV spectrum photodissociate H$_2$O, CO$_2$, and CH$_4$ 
in the 10$^{-4}$ bar (10 N m$^{-2}$) and higher layers, producing 
atomic hydrogen and 
oxygen. EUV radiation below 91.2~nm ionizes hydrogen that then heats 
and inflates the outer atmospheres of close-in exoplanets 
\citep[e.g.,][]{Murray-Clay09,Chadney15}. This is 
important for the evolution of planetary atmospheres, because 
hydrogen is readily lost through hydrodynamic outflow and ion 
pickup by the magnetic stellar wind in distended atmospheres. 
This process may also explain the loss of the Martian atmosphere 
\citep{Brain10}. Whether or not an exoplanet in the habitable 
zone can support life forms depends crucially on the stability and 
chemical composition of its atmosphere. 

The objective of this paper is to create the first complete
physical model of a representative M dwarf, extending from its 
photosphere to its corona, that matches the observed spectrum and can
predict the unobserved portions of the stellar spectrum. 
Because the physical mechanisms that heat the chromospheres 
and coronae of late-type stars are not well understood, we use 
spectra over a broad range of wavelengths to test semi-empirical thermal 
structures of stellar chromospheres and coronae. By computing 
a physically consistent semi-empirical model, we can determine 
where in the atmosphere each spectral line and continuum is formed
and note differences between where these features are formed in M
dwarf models compared to solar models.
Based on a realistic atmospheric model describing these formation
temperatures and heights, one can predict the input spectrum at the
top of an exoplanet's atmosphere for earlier and later times in the
host star's evolution by perturbing the stellar model to fit the
emission-line fluxes (e.g., Ca~II, H$\alpha$, Mg~II) observed in 
stars with different ages \citep[see][]{Engle09}. 
Understanding the radiative output of host stars and
its effects on habitability is essential for identifying the best
targets for future studies of exoplanet atmospheres.
Stellar models also permit us to compute
the net cooling rates that theoretical models must 
eventually explain by predicting the required heating rates at each height 
and temperature.

Although there are extensive grids of theoretical models of cool star 
photospheres computed assuming convective-radiative equilibrium, for 
example the PHOENIX models and their revisions 
\citep[e.g.,][]{Allard95,Allard01,Hauschildt08,Allard10} and the
MARCS models \citep{Gustafsson08}, these models are 
inadequate for predicting stellar UV spectra. 
\citet{Maldonado15} summarize the present state of photospheric
models of cool stars and provides useful intercomparisons.
Existing photospheric models fail by orders of magnitude to predict the
continuum and line flux below 250~nm emitted by the chromosphere and
corona \citep[cf.][]{Loyd16} where nonthermal heating determines the thermal 
structure. Non-LTE line formation is also needed to explain the observed
line profiles and fluxes. Most observed M, K, and G stars, and even 
F types, have upper chromospheres and transition regions in which 
nonradiative heating, most 
likely by the conversion of magnetic energy to heat, raises 
temperatures far above the 
radiative-convective equilibrium predictions. 
This heating decreases the molecular abundances 
in a way that can be confused with a lower stellar metallicity.
 
Previous chromospheric models for M dwarf stars were usually 
computed to fit only the Ca~II, H$\alpha$, and Na~I D lines 
\citep[e.g.,][]{Mauas97,Houdebine97,Short98,Fuhrmeister05,Houdebine09,
Houdebine10a,Houdebine10c}. 
An exception is the M-dwarf flare models computed by 
\citet{Hawley92} that fit the observed X-ray emission and 
transition-region lines in addition to the chromospheric lines. 
Although emission 
in the cores of the Ca~II H and K lines indicates the presence of
nonradiative heating in the upper chromosphere, 
these lines are insufficient for predicting the 
temperature structure above 7,000~K and thus 
most emission lines at wavelengths shorter than
300~nm. With a semi-empirical non-LTE model, \citet{Avrett15} fitted 
the UV continuum and line spectrum
of a sunspot, which may have a thermal structure similar to a
relatively inactive M0~V star but with different gravity, 
very strong magnetic fields, and external illumination from the 
normal solar atmosphere. Given these differences between a sunspot and
an M dwarf and the importance
of exoplanet studies, we compute a semi-empirical
model for a representative M dwarf to fit an excellent panchromatic
data set.

We have selected for modeling GJ~832 (HD~204961), a nearby M2~V 
\citep{Maldonado15}
red dwarf star located at a distance, $d = 4.95\pm 0.03$ pc 
\citep{vanLeeuwen07}. The age of this star is not well constrained 
observationally, but there are clues that the star is not young and
perhaps as old as the Sun. Its measured rotational period, $P_{\rm rot}
= 45.7\pm9.3$~days, \citep{Suarez15} and H$\alpha$ absorption rather
than emission indicate a low level of activity. \citet{West15} finds
that stars with such rotational periods are likely members of a
kinematically older population. \citet{Newton16b} finds that the mean
kinematic age of M dwarfs observed in the MEarth program with rotation
periods in the range 10--70 days is 3.1~Gyr, but this result is
uncertain due to the small number of stars in the group.
\citet{France13} showed that the UV spectrum 
of GJ~832, including the Ly-$\alpha$, C~IV, and Mg~II fluxes, is 
consistent with chromospheric and transition-region measurements 
of other intermediate age M dwarfs. Estimates of the stellar 
properties of GJ~832 are mass, $M = 0.45 M_{\odot}$ \citep{Bonfils13}, 
radius, $R = (0.499\pm 0.017)R_{\odot}$ \citep{Houdebine10a}, 
luminosity, $L = 0.026L_{\odot}$ \citep{Bonfils13}, and  
surface gravity, log g = 4.7 \citep{Schiavon97}.
For these values of the stellar radius and distance, the solid angle
of the star observed at its distance from Earth is
$\Omega_*=\pi(R/d)^2=1.622\times10^{-17}$ steradians, and the solid
angle of the star viewed at a distance of 1 Astronomical Unit (1~AU) is
 $\Omega_{\rm 1AU}=\pi(R/1 AU)^2=1.691\times10^{-5}$ steradians. We will
use these quantities later in the paper to convert from average disk
intensity to flux units and compare with solar models.\footnote{The
stellar radius may have a large uncertainty. If $R$ is as large as
$0.52R_{\odot}$ (see Section 3.1), then $\Omega_*$ and $\Omega_{1AU}$
increase by 8.6\%.}

GJ~832 hosts a $0.64M_{\rm Jup}$ exoplanet \citep{Bailey09,Wittenmyer14}
with a relatively long orbital 
period, $P = 10.01 \pm 0.28$~years, at a mean distance from the star, 
$a = 3.56 \pm 0.28$~AU. \citet{Wittenmyer14} discovered 
that this star also hosts a super-Earth at a mean distance, 
$a = 0.162 \pm 0.017$~AU, with an orbital period, 
$P = 35.67_{-0.12}^{+0.15}$~days. This exoplanet, lies just inside 
the inner edge of the habitable zone, but given its large mass, 
$M > 5.4$ $M_{\rm Earth}$ and likely large greenhouse atmosphere, 
\citet{Wittenmyer14} suggested that it is more likely a very hot super-Venus. 

Although it is difficult to obtain accurate metallicities for M dwarfs,
\citet{Bailey09} derived [Fe/H] = $-0.31\pm 0.2$ for GJ~832 using the 
photometric metalicity calibration of \citet{Bonfils05}. Subsequently,
\citet{Johnson09} estimated the metallicity to be 
slightly subsolar ([Fe/H] = --0.12) by using a grid of 
convective-radiative LTE atmospheric models to fit the infrared 
FeH feature. In Section 3.2, we examine 
the issue of metal abundances of this star in light of the new 
physical model with a warm upper chromosphere. 

In Section 2, we describe our approach and the resulting physical 
model. In Sections~3 and 4, we compare the computed 
spectrum with observations, and in Section~5 we compute the contribution 
functions for the important spectral lines and radiative losses 
as a function of wavelength and temperature as a guide to future 
theoretical heating models. In this section, we also discuss 
where spectral lines are formed in the M-dwarf model and the
similarities and differences between the M-dwarf model and solar models.
Section~6 is our summary of this investigation and our conclusions.






\section{PROCEDURE FOR CALCULATING NON-LTE ATMOSPHERIC MODELS AND
EMERGENT SPECTRA}

Our approach is to model the temperature versus pressure distribution in the
atmosphere of GJ~832 based on the star's observed spectrum. 
This approach departs from the classical grids 
of stellar atmospheric models based on purely theoretical
considerations of convective-radiative energy transport (e.g., ATLAS9,
MARCS, and PHOENIX) as described and intercompared by
\citet{Bozhinova15}, but is
similar to our approach used for computing semiempirical models of the solar 
atmosphere as described more fully in the next paragraph. 
We begin our calculations with an existing photospheric
model for a star with similar spectral type and luminosity as GJ~832.
We then substantially modify this thermal structure
in the lower chromosphere and extend the temperature-pressure distribution 
from the temperature minimum to the corona
to match the observed spectrum.
Given the complexity of our model atmosphere code needed to properly
compute statistical equilibrium and radiative transfer in very 
many transitions, we assume a one-dimensional geometry. The first
models of M dwarfs with a three-dimensional geometry are now being
computed \citep{Wedemeyer13}, but such models must make approximations
in the radiative transfer physics and completeness that our model does
not make.

We compute stellar spectra in full non-LTE using the 
Solar-Stellar Radiation Physical Modeling (SSRPM) tools. 
These tools are an offspring of the Solar Radiation 
Physical Modeling (SRPM) system developed in the course of five papers:
Paper~I \citep{Fontenla07}, Paper~II \citep{Fontenla09}, Paper~III
\citep{Fontenla11}, Paper~IV \citep{Fontenla14}, and 
Paper~V \citep{Fontenla15}. The SRPM system
computes one-dimensional solar 
atmosphere physical models (distributions of temperature and density
with pressure) in full non-LTE for each of the features observable 
on the solar disk and synthesizes from the models their emitted
radiation. This 
modeling of the entire solar atmosphere from the photosphere to 
the corona produced very high-resolution spectra 
($\lambda/\Delta\lambda \sim10^6$) that we then tested against ground- 
and space-based spectra of the solar disk center at very high 
resolution to refine the models. The models also reproduce 
observations of the Sun as a star, the solar spectral irradiance
(SSI), including the UV spectra observed by a number of spacecraft with good 
absolute flux calibration. The most recent models described in Paper~V
now accurately fit the solar 160--400~nm irradiance that previous models
had great difficulty fitting.
The SRPM tools, therefore, provide a very solid 
platform upon which we added to SSRPM a more extensive calculation of 
molecular formation in LTE that includes molecular sequestration 
of elements and an improved list of the molecular lines that are 
most important for M-dwarfs, e.g., TiO, using data from \citet{Plez98}.

SSRPM calculations currently include 52 atoms and ions in full-NLTE, 
in addition to H, H$^-$ and H$_2$. The effectively thin approximation 
is used for an additional 198 highly ionized species. In total, we consider 
18,538 levels and 435,986 spectral lines produced by atoms and 
ions. Also included are the 20 most-abundant diatomic molecules and 
about 2 million molecular lines. As described in Paper~V, we use new
photodissociation cross-sections of the important species.
Radiation in the lower layers is 
computed in a plane-parallel approximation and the coronal 
radiation in a spherically symmetric approximation. The inclusion 
of optical data (see Section 3.1) allows us to obtain 
consistent models that constrain the photospheric structure and visible 
to IR wavelengths. The inclusion of UV emission lines 
originating in the chromosphere and transition region and  
coronal X-ray data allows us to create a self-consistent physical picture 
of the complete M dwarf atmosphere and spectrum from 0.1~nm to 300~microns. 

Our calculations use an iterative linearization scheme to solve 
the nonlinear molecular LTE equilibrium equations. For each general 
iteration in the present scheme, the linearized simultaneous 
equations for molecule formation are solved in a subiteration block, and 
the full non-LTE level populations and ionization states are solved for all 
neutrals and ions in a separate subiteration block. We normalize 
densities to the elemental abundances, which the present model assumes
to be solar. These elemental abundances are listed in Papers III-V.
We note that coronal abundances may differ from those lower in the
atmosphere, the so-called FIP effect.

Qualitatively, the star's atmospheric model is based on our experience 
in modeling the Sun, but its physical values are very different. The conversion 
of the absolute measurement of the stellar spectrum flux at Earth to 
the stellar average-disk intensity is carried out as in \citet{Fontenla99}
and depends on the accuracy of the  
stellar radius-to-distance ratio. Since the distance to the star 
is known to better than 1\%, the stellar radius dominates the
uncertainty. The color index B--V = 1.52 \citep{Bailey09} points to a 
photospheric temperature $T\sim3600$~K \citep{Bessell95}, while 
\citet{Maldonado15} estimate $T_{\rm eff}=3580\pm68$~K. In the
present study, we assume the value of gravity at $5\times10^4$ cm
s$^{-2}$, and radius, $R=0.499 R_{\odot}$ \citep{Houdebine10a}.

\begin{figure}[h]
\epsscale{0.75}
\plotone{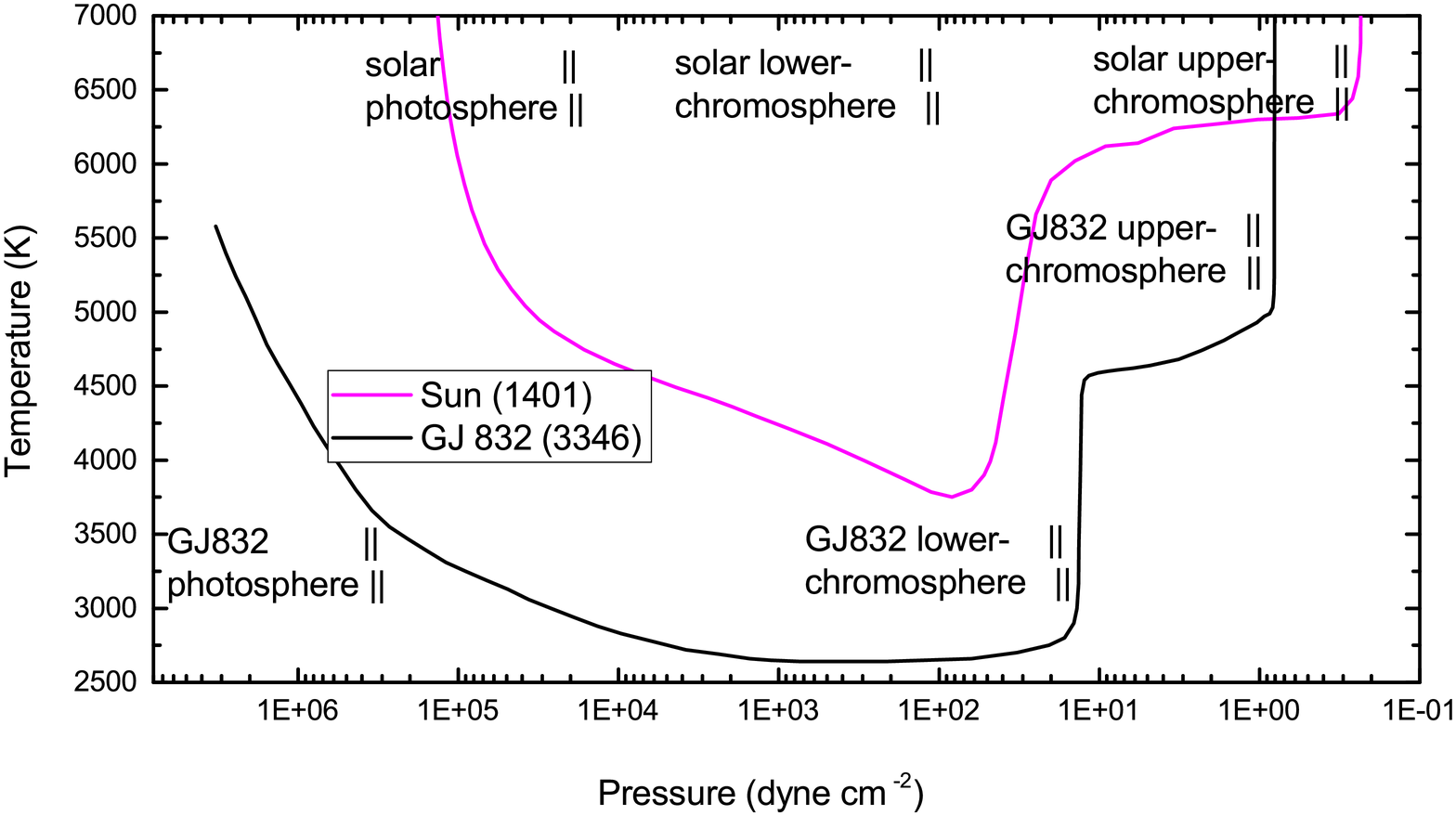}
\plotone{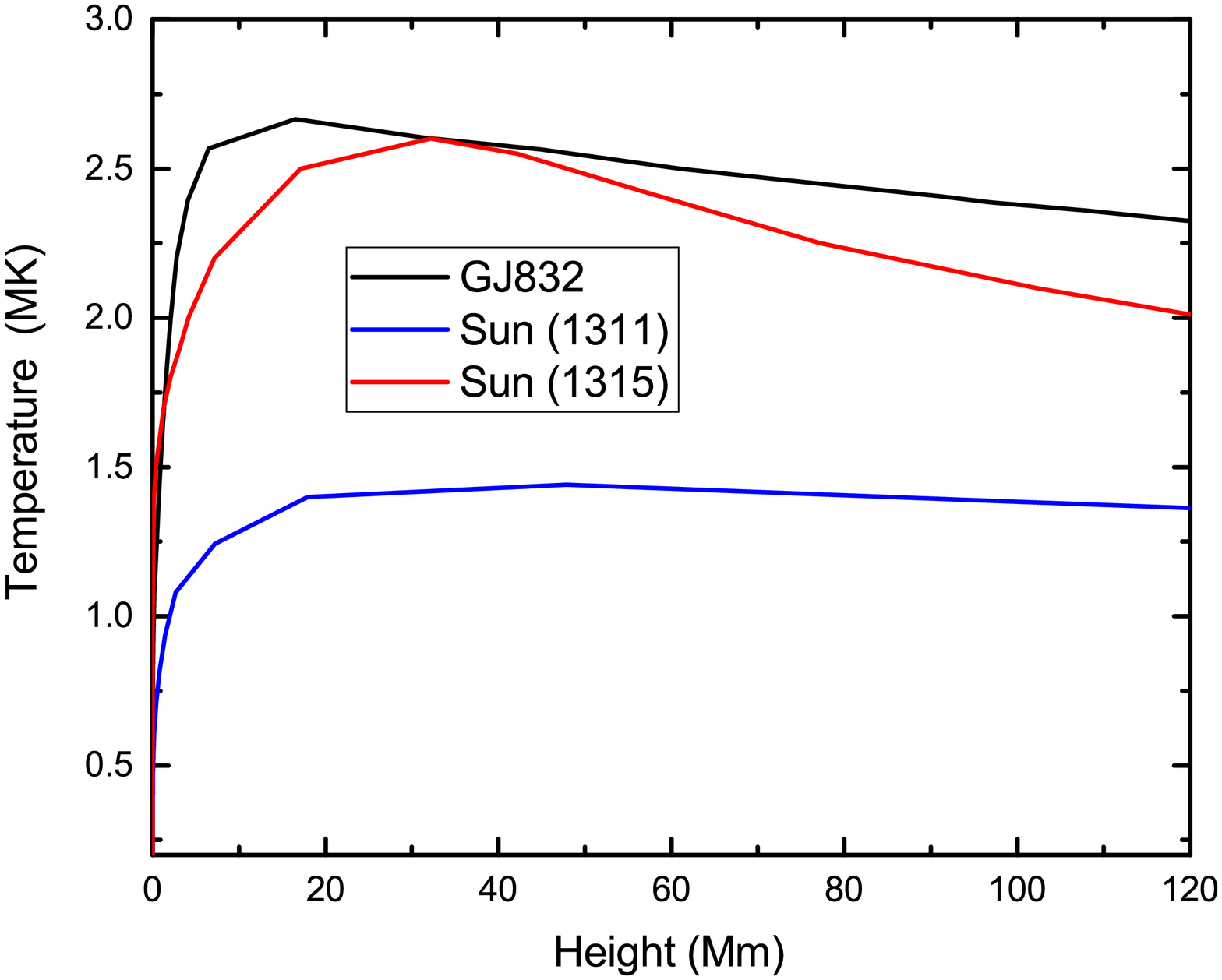}
\caption{{\em Top panel:} Temperature versus gas pressure
distributions for GJ~832 (model 3346, red line) and the 
quiet-Sun inter-network model 1401 (purple line) from Paper~V. 
The double vertical lines designate boundaries: the solar and GJ~832
photospheres occur at higher pressures to the left of 
the left-most sets of double lines, the lower
chromospheres lie between the first and second sets of double lines, 
and the upper chromospheres lie between the second and third sets of 
double lines. The temperature minima of the models are at
$P\approx100$~dynes cm$^{-2}$, and the steeply rising temperature in
the transition region and
extended coronal layers (not shown) lie above the upper chromosphere. 
{\em Bottom panel:} Temperature versus height for the GJ832 model 3346
(black line) compared with the quiet-Sun inter-network model 1311 (blue)
and the very bright solar plage model 1315 (red) from Paper IV.
\label{fig1}}
\end{figure}

In the top panel of Figure~1, we compare the temperature versus gas 
pressure distribution for GJ~832 (model 3346) to that for the quiet Sun  
inter-network model 1401 computed in Paper~V.  
The models are quantitatively very different, but there are many
qualitative similarities. Both models have a photosphere and lower chromosphere
in which temperature decreases outward with decreasing 
pressure. After reaching a minimum, the temperature in the
upper chromosphere rises rapidly beginning at a pressure 
$P\approx15$ dynes cm$^{-2}$ in the GJ~832 model and $P\approx40$
dynes cm$^{-2}$ in the solar model, reaching a plateau
with nearly constant temperature followed by a steep temperature rise
in the transition region. Like solar models, model 3346 includes a 
geometrically thin transition region and a geometrically extended 
corona with a maximum temperature of approximately 2.7~MK (see Section
3.5), which is significantly 
hotter than in models for the quiet-Sun inter-network (model 1311) 
but is similar to the solar corona in the bright plage model 1315
(computed in Paper~IV) as shown in the bottom panel of Figure~1.
We plot  temperatures versus height for the transition region and
coronal layers, since the pressure is nearly constant with height in
these hot layers.

The thermal structure of the atmospheric model of GJ~832 is quantitatively 
very different from the Sun, in particular model 3346 has a much cooler 
photosphere and chromosphere and higher
pressure in the transition region than in the quiet Sun model. As a result,
the spectral features of GJ~832 have different fluxes than the Sun.
The FUV spectra observed by {\em HST} provide important constraints on the 
transition region that we discuss below. 
The detailed structure of the coronal portion of the GJ~832 model 
is mostly theoretical and results from the very high thermal
conductivity of coronal plasma. We use the total X-ray luminosity
and the Fe~XII 124.2~nm line to constrain the energy deposited in the
corona and the energy balance calculations to determine the
temperature structure of the corona of GJ~832.

The photospheric and lower-chromospheric layers are derived from the 
visible spectrum and do not differ significantly from convective-radiative 
standard models, for example the PHOENIX models. However, 
extending the lower-chromosphere temperatures further out in the 
atmosphere completely fails to reproduce the UV emission lines
of ionized species. The GJ~832 model includes a steep temperature
rise with decreasing pressure begining at
$P\approx15$ dynes~cm$^{-2}$. The pressure at which this 
chromospheric temperature rise occurs was first estimated by considering the 
triggering of the Farley-Buneman instability \citep[see][]{Fontenla07}
and a magnetic field of about 100 Gauss, 
which is representative of typical values for the magnetic field observed 
in the quiet Sun. Above the upper-chromospheric ``plateau'' with nearly
constant temperature, there is a sharp rise to a coronal 
temperature $T\approx2.7\times10^6$~K (see Section 3.5) 
with little change in pressure 
because of the small geometric thickness of the transition region.
The steepness of the
transition region is determined from matching the emission lines in the 
observed 100--200~nm spectrum using the methods described in Paper~II with
detailed calculations of the radiative 
losses by H and all other species (see Section~5.2) based on the recent 
atomic data from CHIANTI 7.1 \citep{Landi13}.

\section{COMPARISON OF OBSERVED AND COMPUTED SPECTRA}


We now compare the observed spectrum of GJ~832 with the spectral synthesis 
of our model 3346. For this we use the 
``average disk intensity'' (units erg s$^{-1}$ cm$^{-2}$ nm$^{-1}$ sr$^{-1}$)
as defined by \citet[][Appendix B]{Fontenla99}, 
which is equivalent to the ``astrophysical flux'' at 
the stellar surface \citep[cf.][]{Mihalas78}. The average disk
intensity is a property of the star's atmosphere (independent of
stellar distance and radius) and 
is converted to the physical flux measured at Earth by a
conversion factor that is the inverse of the stellar angular
diameter, $\Omega_*=\pi(R/d)^2$, where $R$ is the star's radius, and $d$ the 
distance from Earth to the center of the star. Using these values 
as mentioned above, we establish the conversion factor of 
$1/\Omega_*=6.165\times10^{16}$ sr$^{-1}$. The 
conversion between astrophysical flux traditional units and 
SI units is simply: 1 W m$^{-2}$ nm$^{-1}$ sr$^{-1}$ = 100 erg
cm$^{-2}$ s$^{-1}$ \AA$^{-1}$ sr$^{-1}$.

\subsection{The Visible Spectrum}

We analyze the optical spectra of GJ~832 observed  on Sept 1 and 15, 2012 
with the REOSC cross-dispersed echelle spectrograph on the 2.15~m 
Jorge Sahade telescope at the Complejo Astronomico El Leoncito
(CASLEO) in Argentina. 
The REOSC spectrum was flux calibrated by assuming that it 
is similar to that of a reference M1.5~V star (GJ~1), which was 
calibrated according to the procedure developed by \citet{Cincunegui04}.
The calibration procedure in that paper was estimated to have 
a 10\% uncertainty but it is not well established how GJ~832 may
differ from GJ~1, implying that the absolute flux calibration
uncertainty may be larger than 10\%.

The observed visual spectrum wavelengths have been converted to vacuum
wavelengths for comparison with the model spectrum, which was smoothed
to the resolution of the CASLEO REOSC spectrograph of 
$\lambda/\Delta\lambda = 13,000$ (2 pixels). 
\citet{Evans61} published a
radial velocity of 4.1 km~s$^{-1}$ for this star, but 
the moderate resolution of the CASLEO REOSC spectrograph 
does not allow for a very accurate wavelength scale and we have not applied
this small Doppler shift to the observations.

\begin{figure}[h]
\plotone{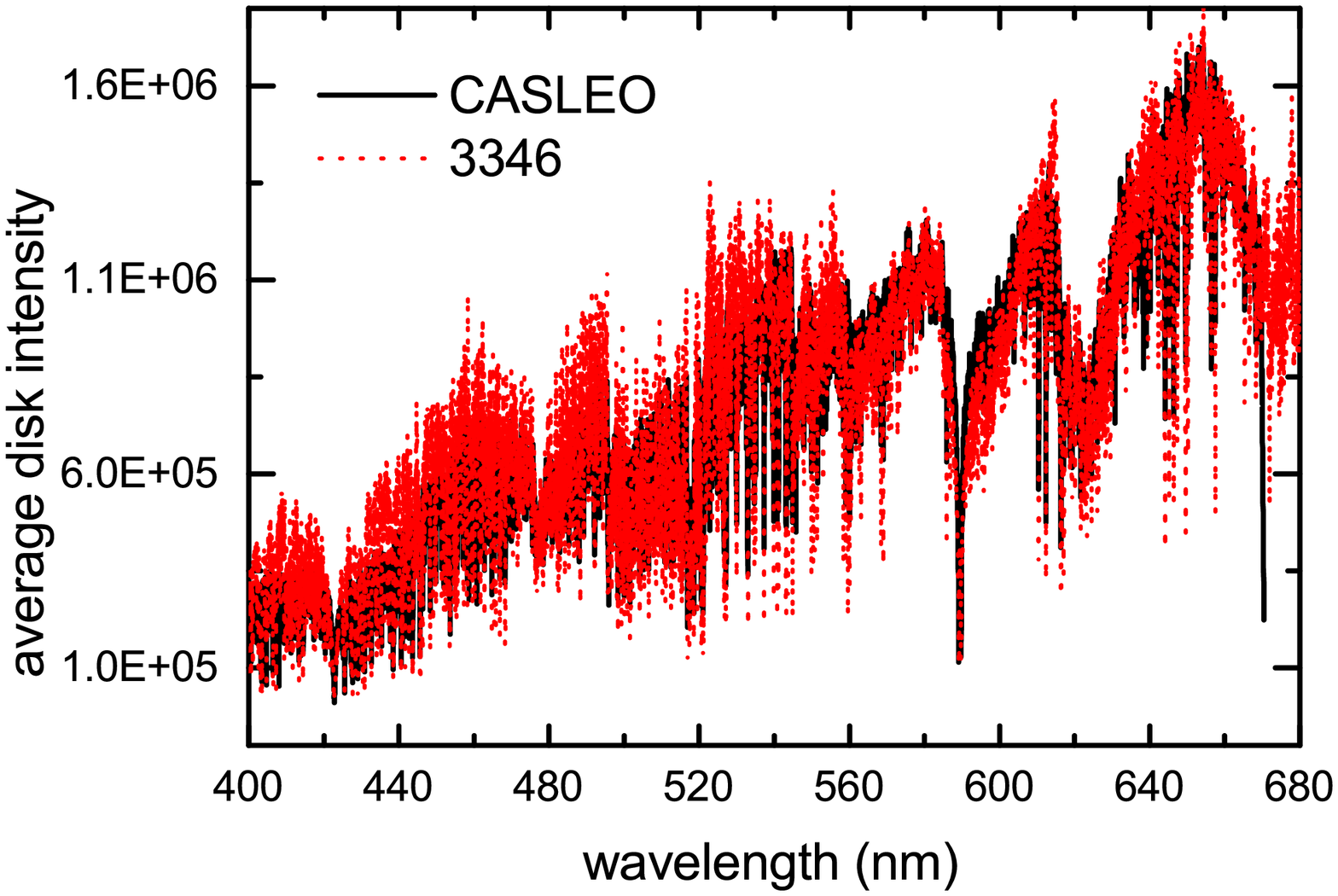}
\caption{The average disk intensity (erg s$^{-1}$ cm$^{-2}$ nm$^{-1}$ sr$^{-1}$)
of GJ~832 in the optical region 
observed by CASLEO on Sep 2012 (black) is compared with the computed 
spectrum from our model 3346 (red).\label{fig2}}
\end{figure}

The observed and computed spectra between 400~nm and 675~nm are shown
in Figure 2. This figure shows 
that the adopted set of parameters and physical model are broadly 
consistent with the observed intensities, although some minor 
differences occur.

\begin{figure}[h]
\epsscale{1.1}
\plottwo{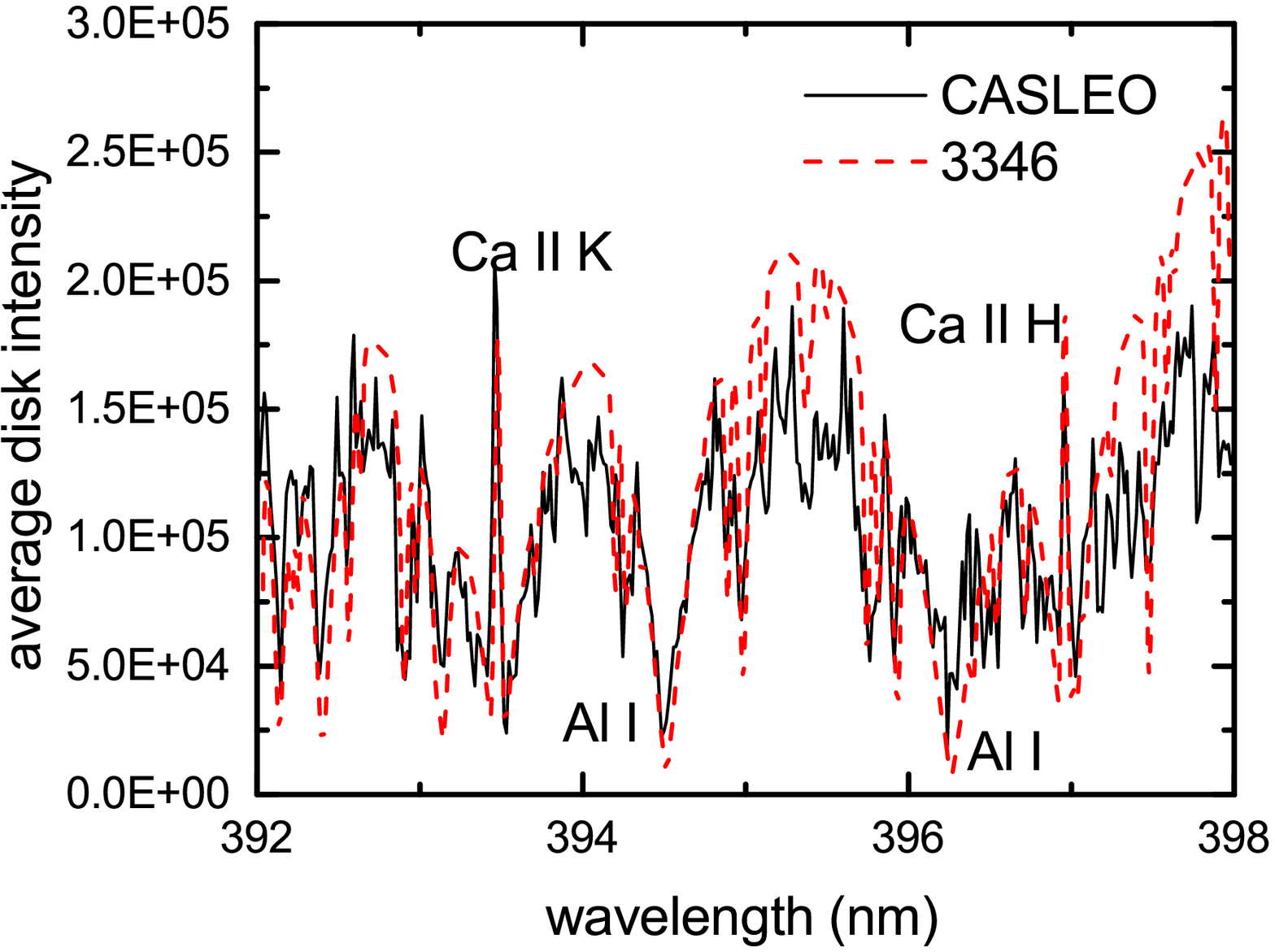}{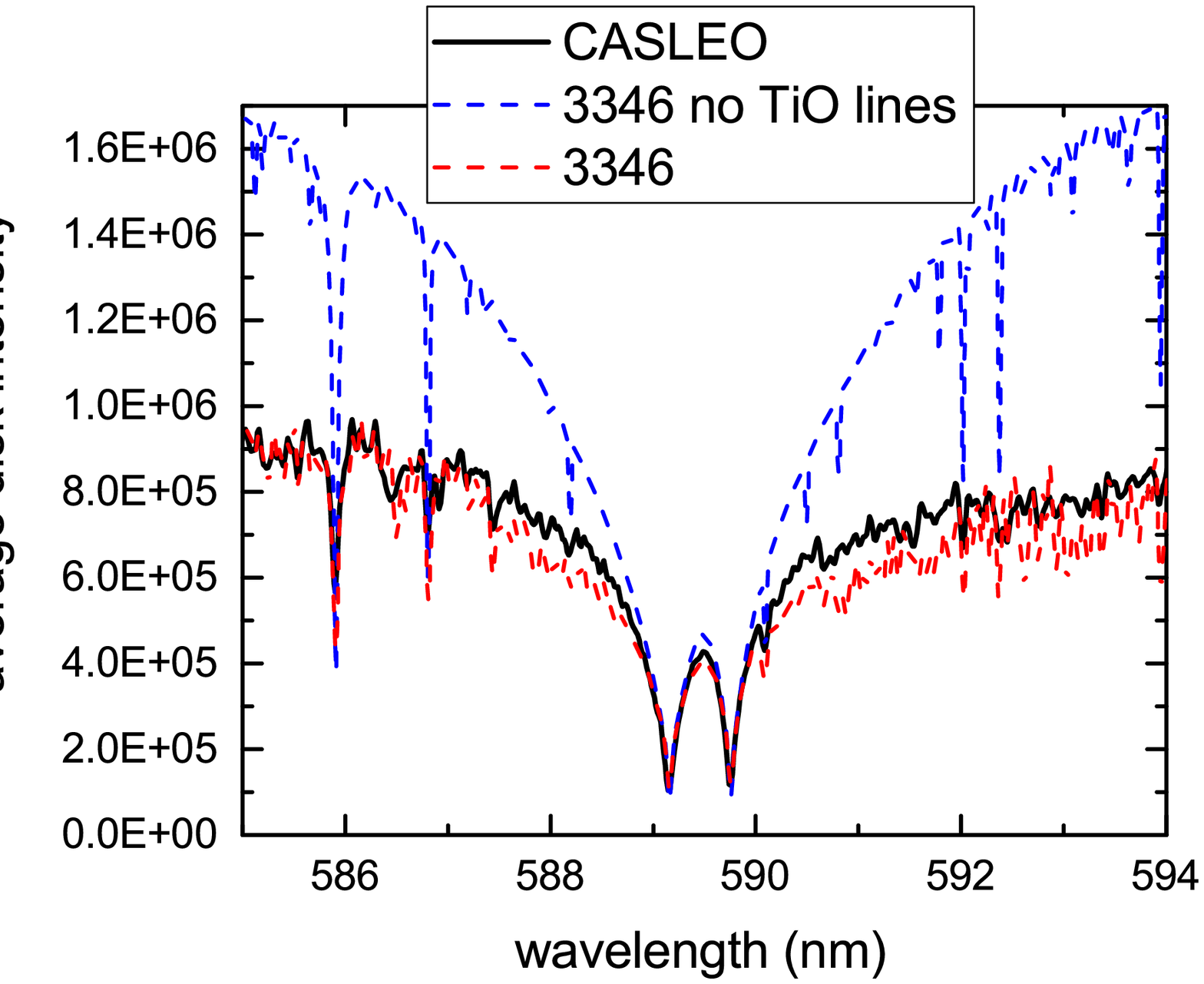}
\epsscale{1.05}
\plottwo{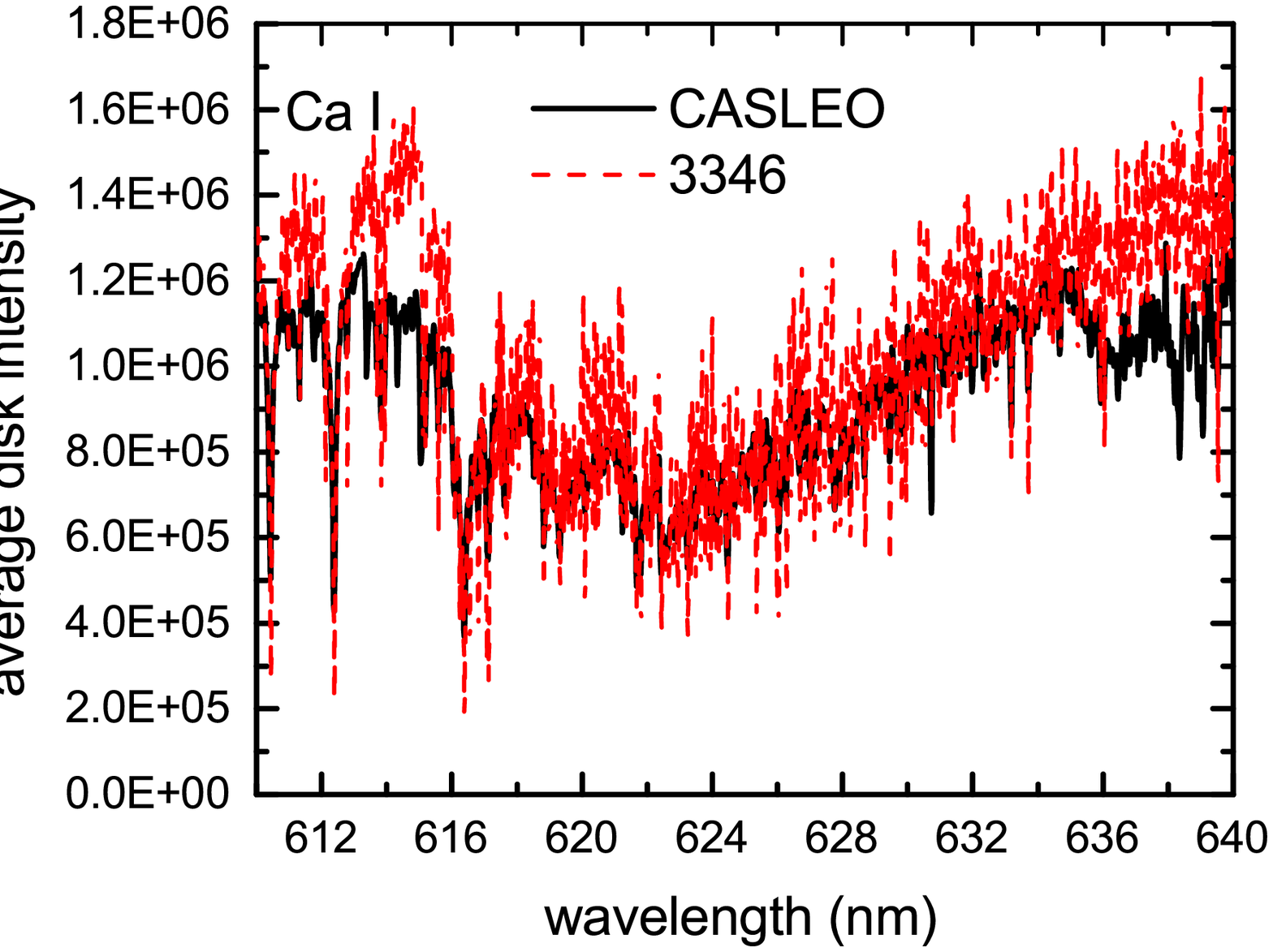}{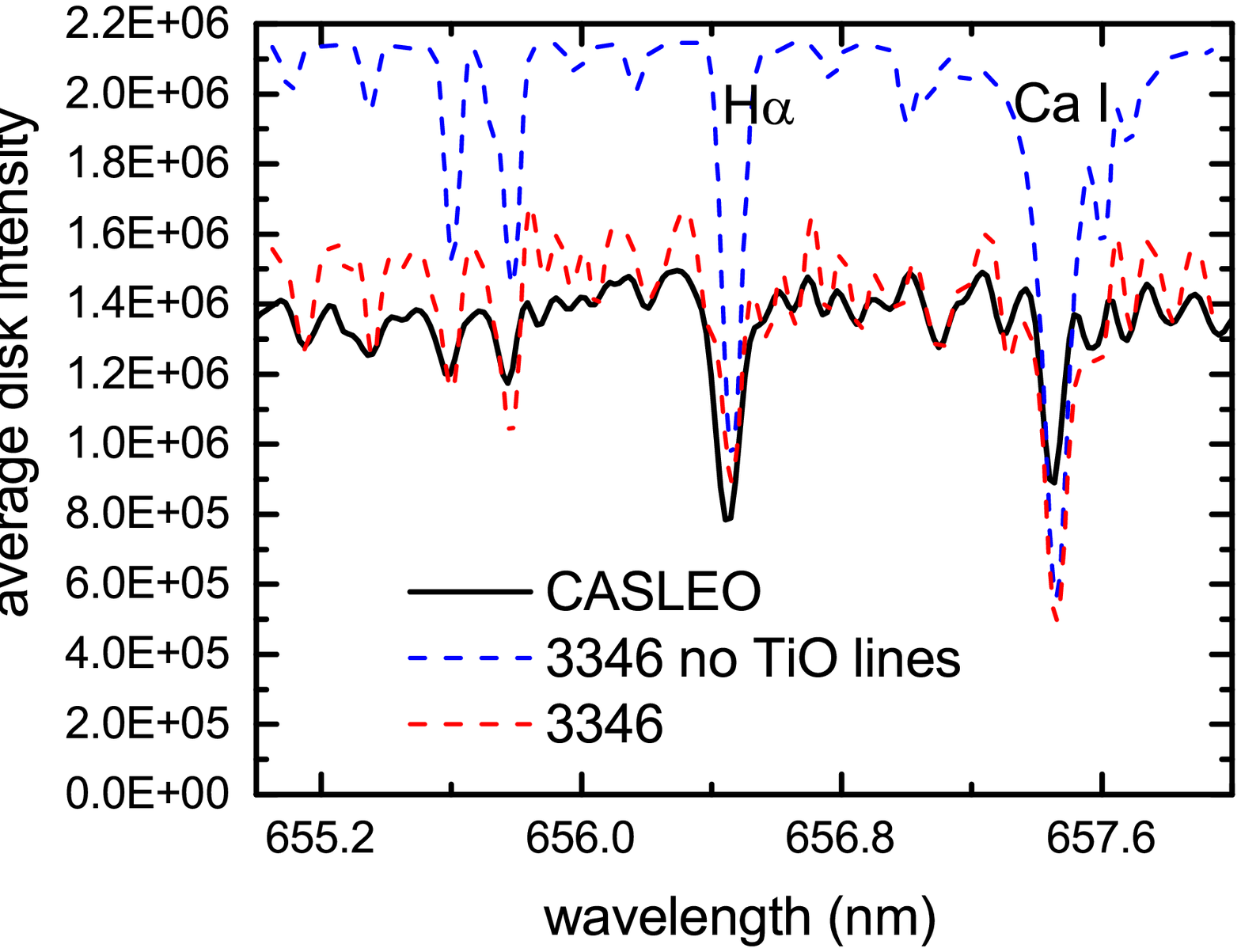}
\caption{Comparison of the average disk intensity 
(erg s$^{-1}$ cm$^{-2}$ nm$^{-1}$ sr$^{-1}$) of GJ~832 observed
at CASLEO (black lines) with the computed spectrum of model 3346 (red lines). 
{\em Upper left panel:} Ca~II H and K spectral region; {\em upper
right panel:} Na~I D lines region; {\em lower left panel:} one of the 
TiO bands; {\em lower right panel:} H$\alpha$ region. The Ca~II H emission 
line is located just below the H symbol in the upper left panel.
The blue lines in the upper right and lower right panels show the
effect of not including TiO line absorption.\label{fig3}}
\end{figure}

\subsubsection{Continuum Intensity}

Our experiments varying the photospheric temperature distribution 
show that the overall shape of the visible spectrum 
depends very little on the photospheric temperature gradient but much 
more on the temperatures at photospheric pressures. For solar-type 
stars, the layer at which optical depth is unity in the radial
direction at the continuum wavelength of 500~nm is often used as 
a reference height. This reference height is not practical for M-type 
stars, because many spectral lines  
and the continuum cannot be isolated in medium-resolution spectra such
as those obtained with the CASLEO REOSC. 

While optical depth unity ($\tau =1$) is a representative height for
the observed photospheric emission at solar disk center,
$\tau=2/3$ corresponds to optical depth unity at the mean observing 
angle ($\mu=2/3$) for an unresolved star and is more 
representative of the average disk intensity or emitted flux. We 
therefore consider $\tau=2/3$ as the approximate location for stellar emission.
The observed pseudo-continuum is much lower than the theoretical real
continuum because of the ``jungle" of overlaping molecular absorption 
lines, mostly TiO lines, that cannot be separated even in
high-resolution spectra. The molecular line features that are discernable in
the CASLEO spectrum correspond to exceptionally high transition 
probabilities, and are
less well defined in the observed spectrum than in our computated spectrum,
because we include only micro-turbulent line
broadening and stellar rotation velocities. Other velocities would
smear the model spectrum but do not affect the conclusions of the present paper.
For instance, in
our high-resolution spectrum near 650~nm, the highest flux 
occurs at $\tau=2/3$, where the pressure 
$P_{\rm  phot}\sim1.8\times10^5$ dyne cm$^{-2}$ and the temperature 
$T_{\rm phot}\sim3440$~K. However, at the bottom of the nearby spectral lines,
$\tau=2/3$ occurs higher in the atmosphere at $P\approx2.6\times10^4$ 
dyne cm$^{-2}$ and $T\approx3000$~K. In a calculation that excludes molecular 
lines but uses the same physical model, we find that the true continuum 
formed near $\tau=2/3$ occurs at 
$P_{\rm phot}\sim2.8\times10^5$ dyne cm$^{-2}$ and 
$T_{\rm phot}\sim3570$~K.
However, the highest intensity level between spectral
lines is significantly below the true continuum, which cannot be observed 
in the visible spectrum even at extremely high spectral resolution 
because of line blending.

\subsubsection{Ca II H and K, Na I D, TiO lines, and H$\alpha$}

Figure 3 shows spectral lines from several atoms, ions, and TiO
molecules. Particularly interesting are the visible lines that can 
display stellar activity, including 
Ca~II H and K and H$\alpha$.
Although relatively weak in less active M dwarfs, these lines in
GJ~832 indicate the presence of a nonradiatively heated upper 
chromosphere and are, therefore, valuable proxies 
for the UV flux and stellar activity cycles in M dwarfs.
The upper left panel in Figure 3 shows that the observed 
Ca~II lines are well matched by our model. The computed
line centers of the Ca~II H and K lines form in the
upper chromosphere (see Section 5.1) in the temperature range 
4,000--20,000~K and pressure range 1--10 dynes cm$^{-2}$,
which is similar to the formation temperature range 
in the quiet Sun. In the GJ~832 model, Ca is essentially neutral throughout 
the cool lower chromosphere 
and, as a result, the Ca~II lines do not have the very broad
absorption wings that are formed in the Sun's much warmer lower
chromosphere (see Fig. 3). In the upper chromosphere, Ca~II is 
sufficiently abundant to produce observable 
emission cores but no significant absorption wings. 
Therefore, the Ca II lines in GJ 832 are very different from the
solar spectrum where these lines have
very broad absorption wings but only weak central emissions.

Among other
absorption lines also displayed in this panel, the deepest are 
the Al~I resonance lines at 394.512 and 396.264~nm that are also 
very prominent in solar spectra. The central intensities and widths 
of the Al~I lines are 
well fit by our model, and these lines are much broader than 
observed in solar spectra, because Al is primarily neutral in the
photosphere and lower chromosphere of GJ~832.

Other lines from neutral species, for instance the Na~I D1 and D2
lines, are prominent absorption lines formed in the 
lower chromosphere. The upper right panel in Figure~3 shows that 
the observed and computed cores and wings of the Na~I lines are 
in excellent agreement. The plot also shows that not including
absorption by the many TiO lines would produce a very poor fit to the
Na~I line wings and adjacent continuum.  
Another important absorption line in this panel is the Ca~I 585.907~nm 
line that our model fits very well.

The visible spectrum of GJ~832 is largely dominated by TiO 
absorption lines, as is typical for M-type stars and, to a lesser
extent, for sunspots. These and 
other molecular lines distort the shape of the underlying continuum.
The lower left panel in Figure 3 shows a portion of the important TiO 
bands, indicating that our model produces good agreement in this particular
band. We cannot, therefore, make a statement concerning the metal 
abundances based only on the TiO lines, but we can say that they are 
not inconsistent with the solar abundances we used.

The lower right panel in Figure~3 shows that the observed 
and computed H$\alpha$ line depths are in excellent agreement only when
the many TiO lines are included.
An analysis of the H$\alpha$ line formation indicates that the line
is formed essentially in the upper chromosphere where
the first-excited level population of hydrogen 
produces the H$\alpha$ absorption. This panel also shows the 
strong Ca~I 657.459~nm line whose line center is underestimated,
likely due to poor atomic data for some of the Ca I lines.
  
\subsection{Molecular Formation and Metallicity}

Our computed spectrum is based on the solar abundances listed in
Paper~V, and the relatively small 
differences between the computed and observed TiO lines rule 
out a large difference between the solar and stellar abundances 
of Ti and O. If the abundances of these elements in GJ~832 were substantially 
smaller than solar, then the general slope of the spectra and 
the depth of the TiO bands would not be as close to 
the observations as our computations show. Also, significantly subsolar
abundances would make it difficult to match the observed 
Ca~II, Mg~II, and many other lines.

The inclusion of a chromosphere in
our model leads to a good fit to many, but not all, of the observed
spectral features of GJ~832. We find that the 
metal abundances in this star are consistent with or slightly lower 
than solar, for example log[Fe/H]=--0.12 \citep{Johnson09} and
$-0.17\pm0.17$ \citep{Maldonado15}. \citet{Johnson09}  
argue that the photometric-based metallicity determinations for M dwarfs,
e.g., log[Fe/H] = --0.31 for GJ~832 \citep{Bailey09}, systematically
underestimate these metallicities. Also, the abundance estimate by 
\citep{Bailey09} would reduce by a factor of 4 the depths 
of the TiO lines, which depend on the square of the metallicity, 
and would also decrease the absorption in all other spectral lines, 
making it very difficult to match the observed spectra.

\begin{figure}[h]
\epsscale{1.05}
\plotone{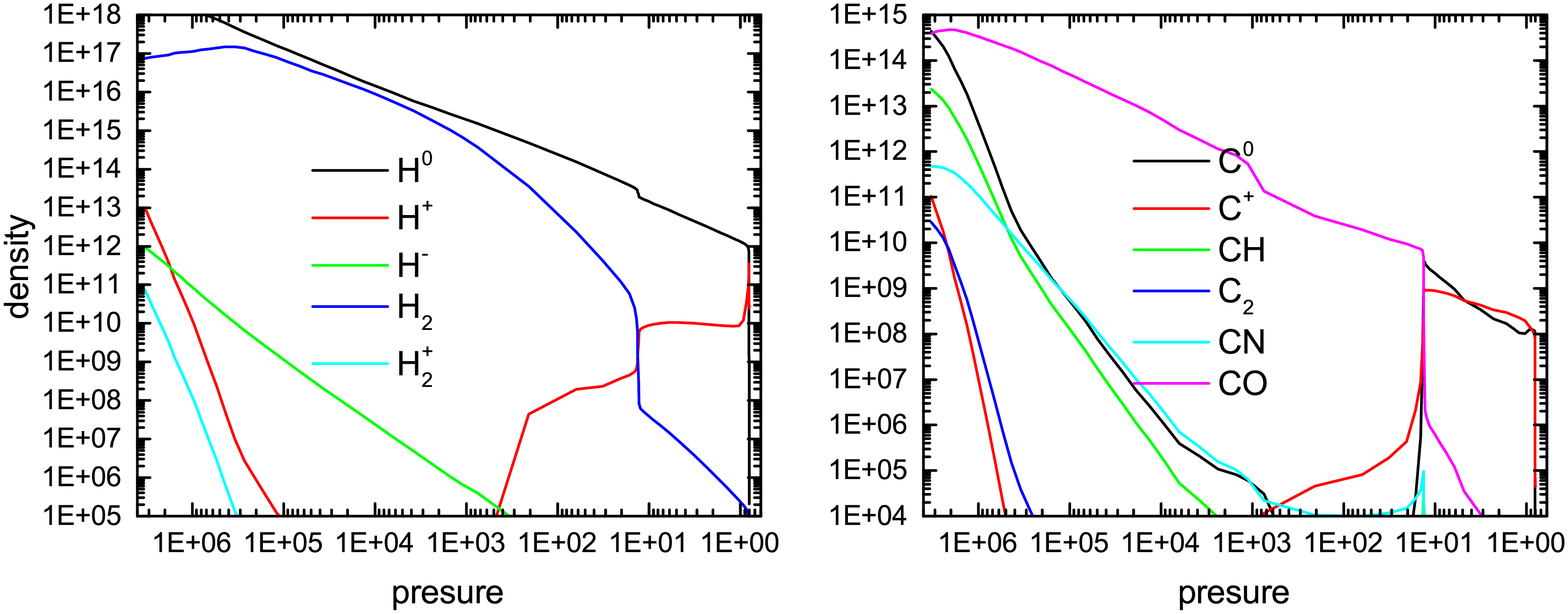}
\caption{Computed densities for H ({\em left panel}) and C ({\em right
panel}) in atomic, ionic, and molecular 
forms. In the photosphere, H$_2$ is as abundant as atomic H, and CO
is the dominant form of carbon below the upper chromosphere. 
In the upper chromosphere where the
temperature rises steeply at $P\approx1$ dyne cm$^{-2}$, singly ionized 
H and C both become the dominant species.\label{fig4}}
\end{figure}

Figure 4 shows some molecular and atomic densities computed in the 
present model of GJ~832. For the elements shown here, the ionization 
remains very low in the photosphere and lower chromosphere,
facilitating the formation of molecules from neutrals.
The left panel in Figure~4 shows that atomic and molecular H have nearly
equal abundance throughout the photosphere. However, the 
H$_2$ density decreases rapidly as the temperature increases and 
density decreases toward the upper chromosphere. In these layers 
ionized H dominates over H$_2$. Despite the temperature 
increase, H remains mostly neutral in the lower and 
upper chromosphere. 
Unlike the Sun, H ionization in GJ~832 remains small even 
in the upper chromosphere, and its contribution to the electron 
density is much smaller than in the Sun.
The right panel of Figure~4 shows C densities in several forms.  
Notably throughout the photosphere and lower chromosphere, CO dominates, 
leaving very little C in other forms. With increasing temperature and 
decreasing density in the upper chromosphere, the CO density decreases
rapidly and 
neutral C atoms dominate, except close to the top of the 
upper chromosphere where ionized C slightly dominates.

\subsection{Near Ultraviolet Spectrum}

\citet{France13} described the NUV spectra of GJ~832 obtained with
the STIS instrument on {\em HST} on 2012 Apr~10 and
2012 Jul~28. The NUV spectrum (157--315~nm) was obtained with the G230L 
grating, which has
low spectral resolution ($\Delta v\approx 600$~km~s$^{-1}$). The Mg~II
lines were also observed at high resolution ($\Delta v\approx
2.6$~km~s$^{-1}$) with the E230H grating that includes the
267--295~nm spectral region.

These data were calibrated to disk average intensity 
using the nominal calibration factor $6.165\times10^{16}$ sr$^{-1}$ 
and Doppler-shifted to match the computed spectrum wavelengths. 
To  match the resolution of the observed spectrum with the computed 
line widths, we convolved the 
computed spectrum with a cos$^2$ filter of 0.3~nm FWHM. 
The faint continuum is
not well measured at wavelengths shorter than $\sim220$~nm, but is 
well detected in the MUSCLES Treasury Survey data \citep{Loyd16}. 

The observed and computed spectra in the 210--300~nm NUV range 
are shown in Figure~5. This figure shows the Mg~II h and k lines 
for which the observation and the model agree well after the observed
fluxes are increased to account 
for interstellar absorption as described below. These 
lines were used to adjust the model's upper chromosphere and 
transition region. We computed the Mg~II and Ca~II lines by merging the
atomic levels of each ion that have the same quantum numbers, 
except for the total angular momentum ($J$), into superlevels and assumed
that the relative populations of the sublevels are in LTE.
This joining together of levels is often used to reduce the size 
of the system of equations while
including many excited states. However, in low density layers, this
approximation is inaccurate, leading to one emission line of the 
multiplet being too bright and the other too faint compared to the 
calulated intensities for the merged state.

In the model for GJ~832, the line centers of the Mg~II lines (at the 
calculation's highest resolution) form near $T=7000$~K at a pressure 
of $P\approx$0.6--0.8 dyne cm$^{-2}$, but the emission peaks on
either side of line center are 
formed mostly at lower temperatures (and higher pressures).  
The Mg~II lines of GJ~832 do not display absorption wings (see
Figure~6), 
because there 
is very little singly ionized Mg in the cool lower chromosphere.
The Mg~II line cores in GJ~832 are narrow because Mg~II is the 
dominant ionization stage only in the upper chromosphere and 
transition region.

The high-resolution Mg~II spectra obtained with the STIS E230H
grating were calibrated to average disk intensity using the 
nominal calibration factor and corrected for a Doppler shift of 12 km s$^{-1}$.
Figure 6 shows the observed and computed high-resolution spectra of 
the Mg~II h and k lines. The computed and observed fluxes are roughly in 
agreement, but the computed profile shapes at line center show a 
double-peaked upper-chromospheric component with a relatively strong 
central reversal. The observed line shape, however, shows only 
a double-peaked profile with a pronounced central 
dip. 

We believe that the observed shapes of the Mg~II
line profiles can be explained as a superposition of a weak-chromospheric
central reversal computed using model 3346 and interstellar Mg~II absorption
at nearly the same wavelength. The \citet{Redfield08} model of the local
interstellar medium predicts that the line of sight
to GJ~832 (radial velocity 4.3 km~s$^{-1}$) likely passes through four
interstellar clouds. According to the kinematic calculator based on this 
model,\footnote{http://lism.wesleyan.edu/LISMdynamics.html} the cloud
velocitites for this line of sight are LIC (-11.3 km~s$^{-1}$), 
Aql (+8.5 km~s$^{-1}$),
Mic (-15.1 km~s$^{-1}$), and Vel (-17.0 km~s$^{-1}$). The predicted
absorption by the three clouds at negative radial velocities would
occur at --0.015~nm to --0.020~nm from line center 
and may be detected as weak absorption just to the 
left of the emission line.
Absorption by the Aql cloud should occur at +4.2 km~s$^{-1}$
or 0.004~nm from line center. This interstellar absorption 
could add to the central dip in
the k line and the asymmetry with the blue peak slightly
stronger than the red peak. We think that this combination of the
weak self-reversal predicted by model 3346 and interstellar Mg~II absorption
is the most likely explanation for the Mg~II line profile shapes, 
and to compensate for the interstellar absorption
we increase the observed flux of the Mg~II lines 
listed in Table 1 by 30\% \citep[see][Figure~5]{France13}.
The observed fluxes of the Mg~II and other
spectral lines at a distance of 1~AU from the star are compared in Table~1
with the line fluxes computed for the 
GJ~832 and the quiet and active Sun models. 

\begin{figure}[h]
\plotone{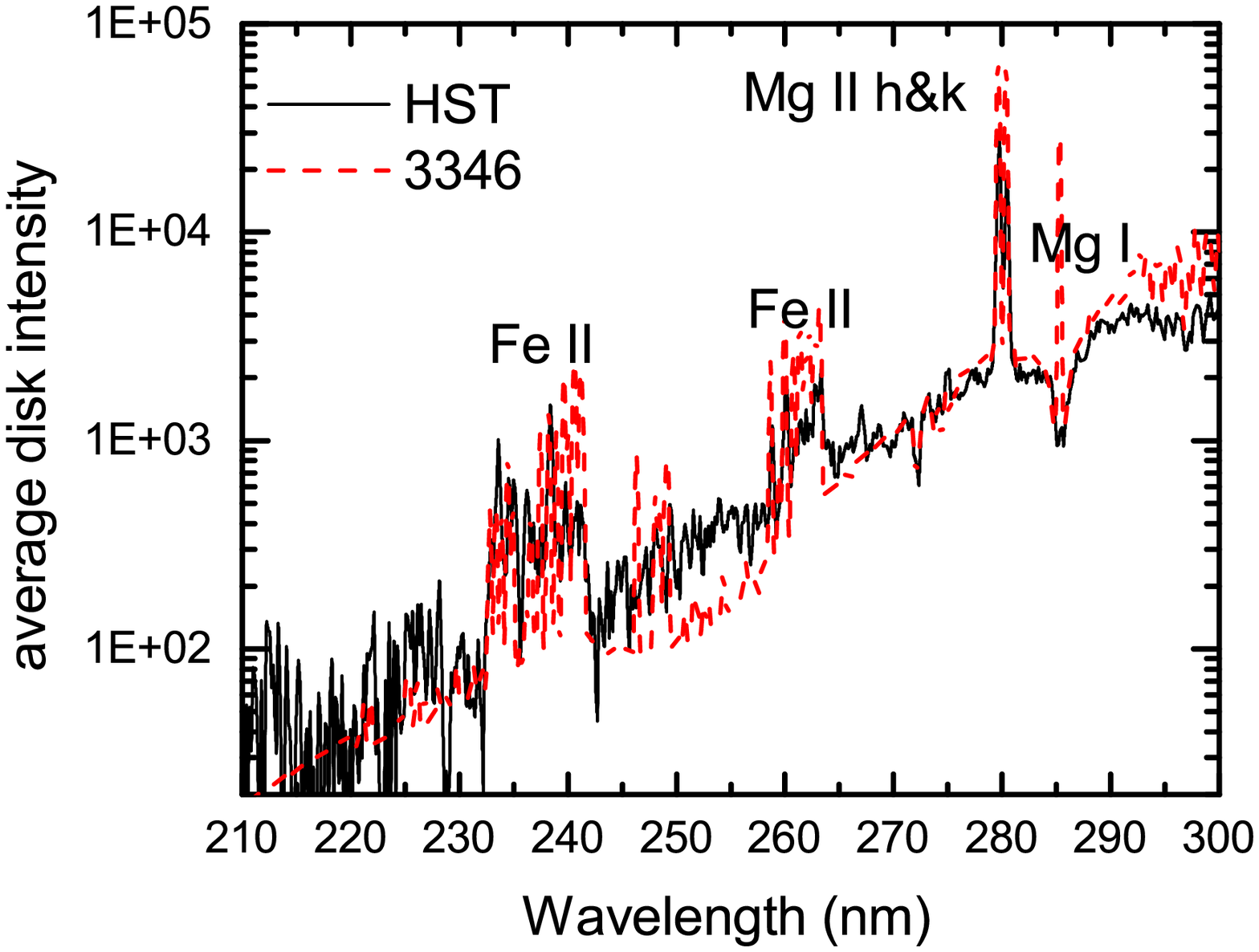}
\caption{Model 3346 (red line) and observed {\em HST} (black line) 
NUV spectra of GJ~832 average disk intensity. Prominent spectral 
features are marked.\label{fig5}}
\end{figure}

Figure 5 shows that the formation of the Mg~I 285.2~nm line
is a problem for the present model as the calculated emission is far
larger than observed. We find that the Mg~I emission forms over the bulk of the 
upper chromosphere from its base pressure $P\approx15$ dyne cm$^{-2}$, 
up to its top pressure $P\approx0.8$ dyne cm$^{-2}$, which produces 
a wide emission line. We found that slight changes in temperature 
and shape of the upper chromosphere can decrease the Mg~I 
emission, since a slight increase in the temperature of this
layer could produce 
a better balance between the Mg~II and Mg~I lines while 
maintaining the good agreement with the observed Ca~II lines.
However, we could not eliminate the strong Mg~I emission, and in Section 5.4
we argue that uncertain atomic ionization rates are the more likely 
explanation for the computed Mg~I emission feature.

Many other lines shown in Figure~5 show good agreement between the calculations
and the observations. Particularly important are the many Fe~II 
emission lines in the ranges 232--242 and 260--265~nm. Emission in these lines 
would have important effects on O$_3$ for any potentially habitable 
exoplanets around GJ~832 \citep[see][]{Tian14}. 

H$_2$ spin-forbidden transitions are likely an important opacity
source in the 240--310~nm region. Unfortunately, there are no
{\em ab initio} calculations for this opacity source, and in Paper~V
we estimated the strength of the H$_2$ opacity needed to best
fit the 160--300~nm quiet-Sun spectrum. However, we find 
that this strength for the H$_2$ opacity is too large for GJ~832 that has 
lower temperatures than the quiet-Sun model 
1401. Different assumptions regarding the H$_2$ 
opacity greatly affect the NUV continuum intensity in the range 
245--310~nm.
However, the required amount of H$_2$ opacity is closely connected
with the plausible NLTE formation of H$_2$ near the temperature 
minimum, which we will address in a future paper.


\begin{figure}[h]
\plotone{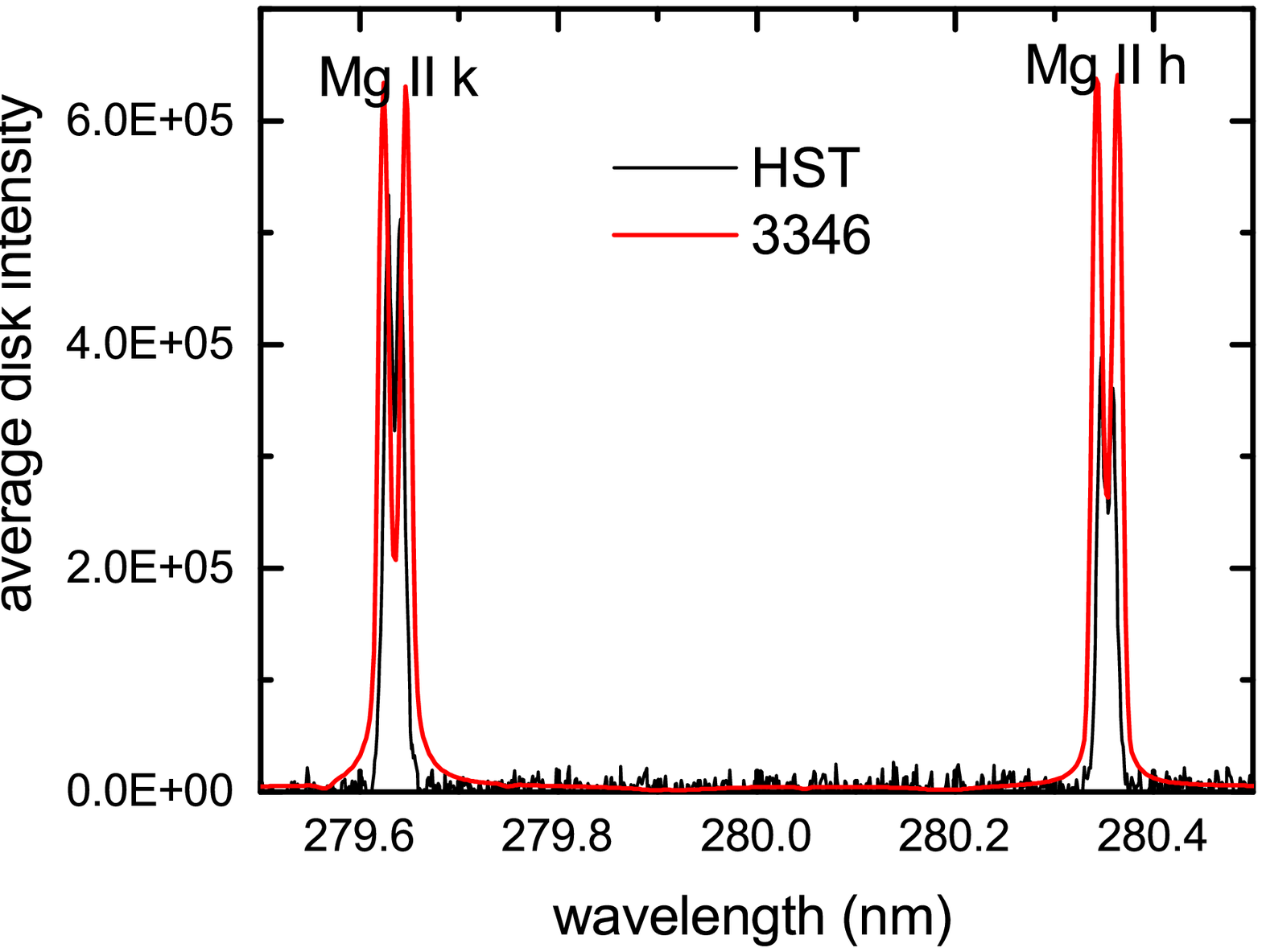}
\caption{Comparison of high-resolution intensity (erg cm$^{-2}$
  s$^{-1}$ \AA$^{-1}$ sr$^{-1}$) of the Mg~II h 
(280.3~nm) and k (279.6~nm)
lines of GJ~832 computed from model 3346 (red line) and observed
(black line) with {\em HST}. The computed Mg~II line profiles do not
include interstellar absorption (see Section 3.3).\label{fig6}}
\end{figure}

\subsection{The Far-ultraviolet Spectrum}

\citet{France13} described the FUV spectra of GJ~832 obtained with
the COS and STIS instruments on 2011 Jun 9
and 2012 Jul~28. The COS G130M and G160M gratings observed 
the 114--179~nm spectral region
with a resolution $\Delta v\approx 17$~km~s$^{-1}$, and the STIS  
E140M grating observed the 114--171~nm spectral resolution with a 
resolution $\Delta v\approx 7.5$~km~s$^{-1}$. We find very good
agreement between the calculated and observed FUV emission lines shown
in Figure~7.

\begin{figure}[h]
\plottwo{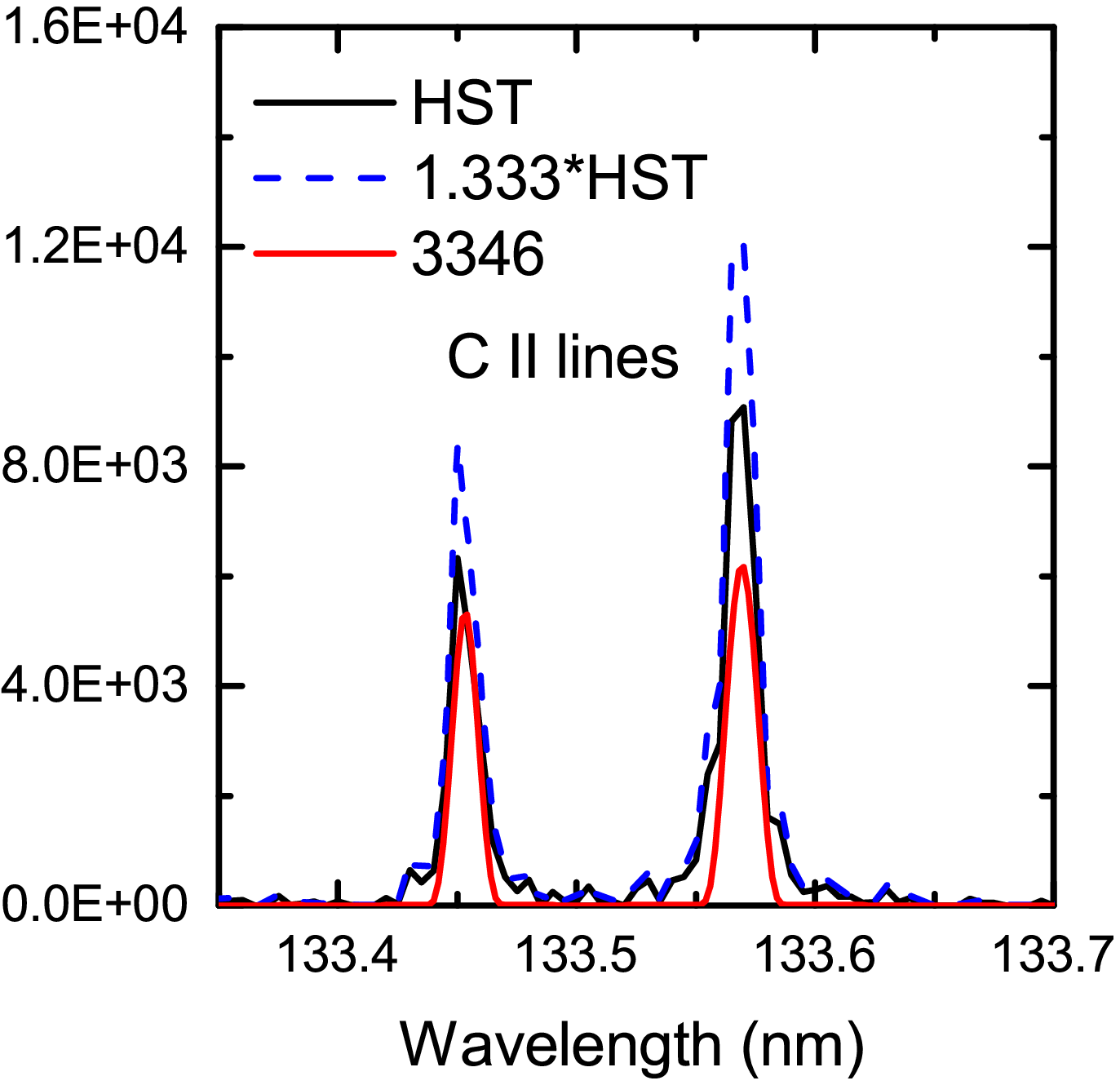}{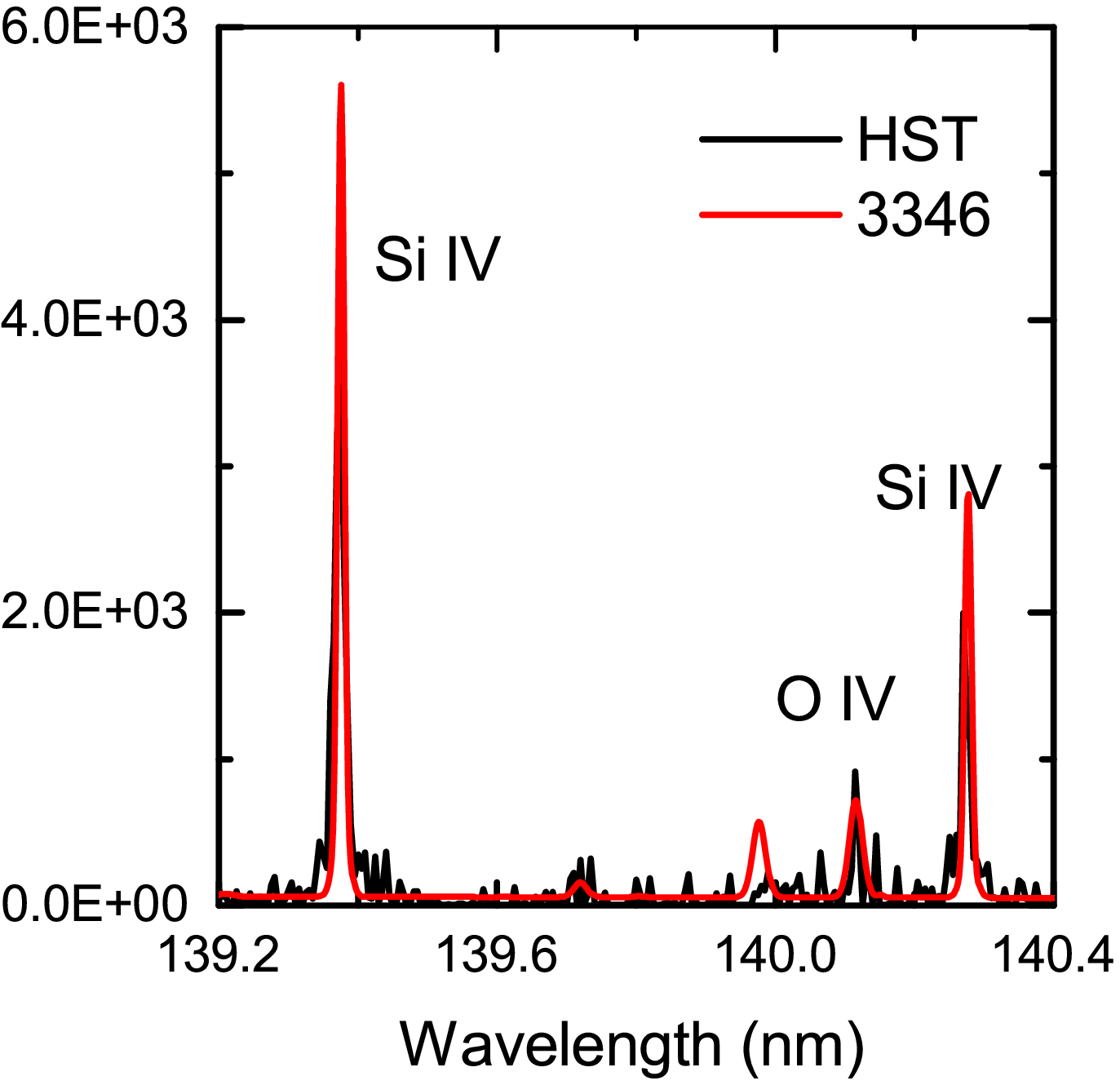}
\epsscale{1.05}
\plottwo{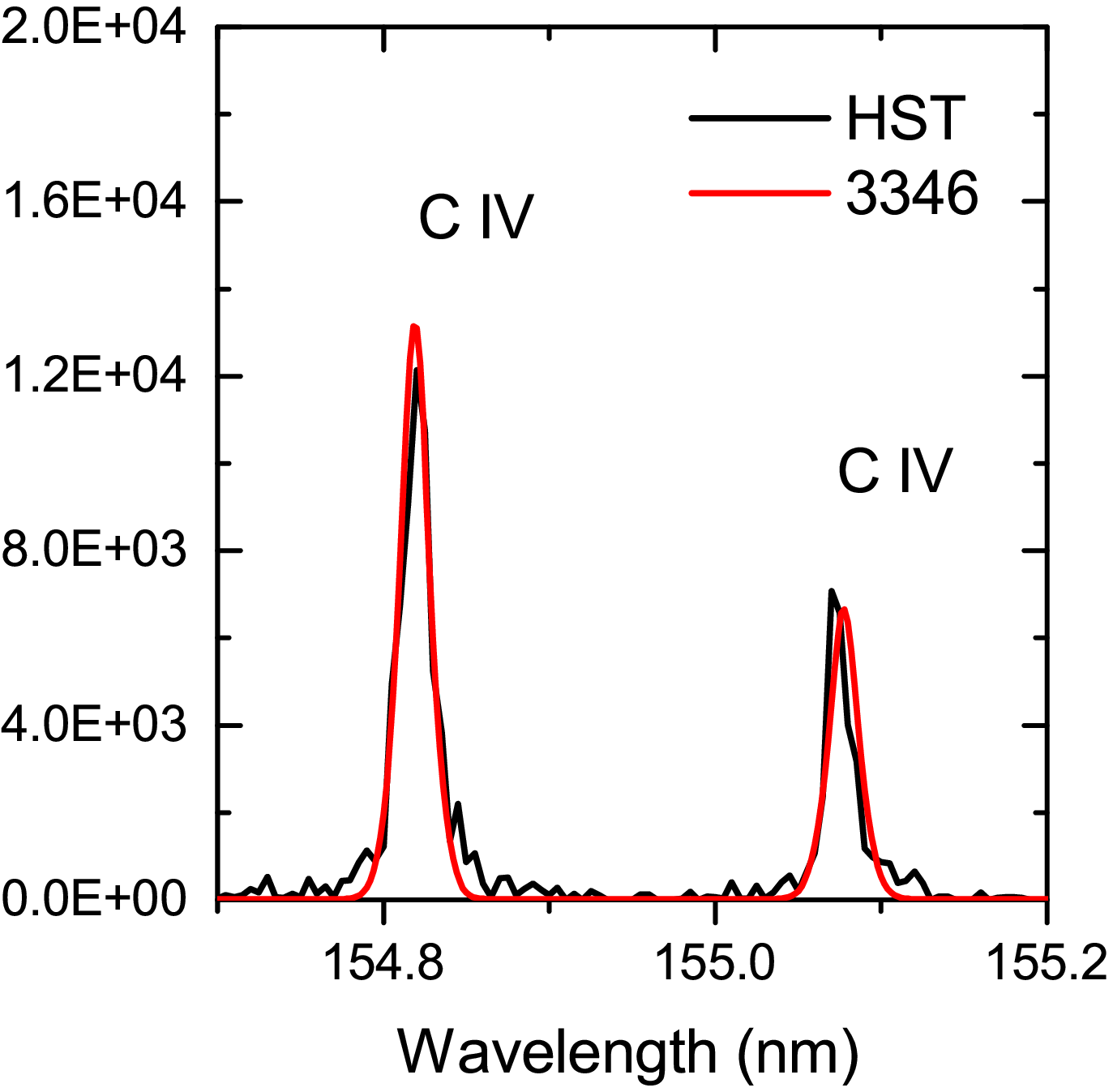}{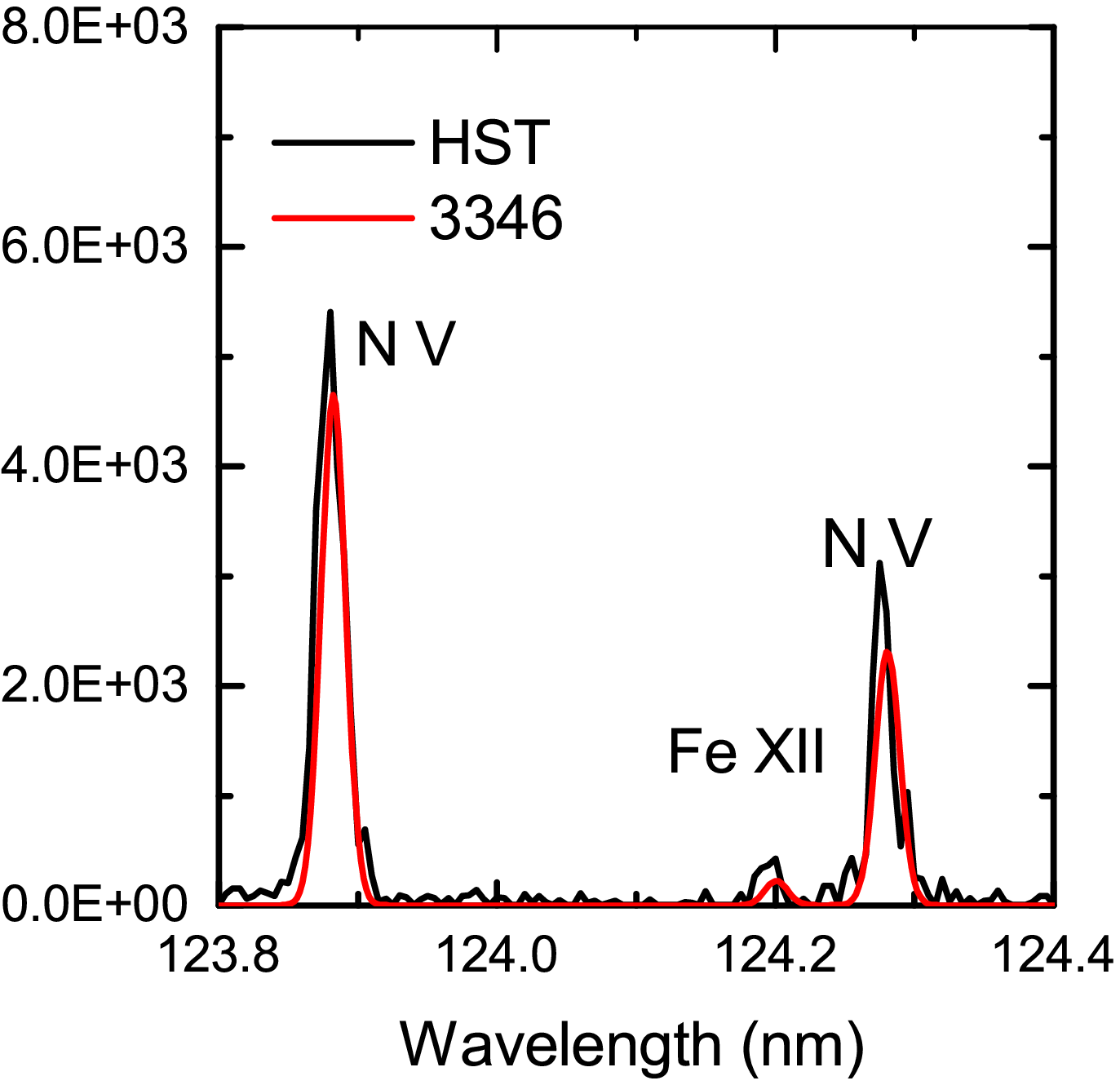}
\caption{Comparison of model 3346 average disk intensity UV line 
profiles (red lines) with GJ~832 line profiles (black lines) observed by 
{\em  HST}. {\em Upper left:} C~II line centers are formed near
$T=25,000$~K. The dashed blue line shows the 30\% correction to the
observed intensity for interstellar absorption.
 {\em Upper right:} Si~IV lines are
formed near 50,000~K and O~IV lines are formed near 160,000~K; 
{\em Lower left:} C~IV lines are formed near
100,000~K; and {\em Lower right:} N~V lines are formed near 170,000~K and
the Fe~XII line is formed near $1.4\times10^6$~K.\label{fig7}}
\end{figure}

In our model, the peak in the contribution function for the 
C~II 133.4~nm line center is near 20,000~K (see Section 5.1), and the
peak contibution function for the Si~IV 139.3, 140.3~nm doublet is 
near 40,000~K. The O~IV] semi-forbidden 140.1~nm line, which
forms at temperatures around 160,000~K, is well fit 
by our model. The C~IV 154.8, 155.1~nm
doublet peak contribution is at 100,000 K and the N~V 123.9, 124.3~nm doublet 
peak is at 170,000~K, but both form over a wide range in temperature.

\begin{figure}[h]
\epsscale{1.1}
\plotone{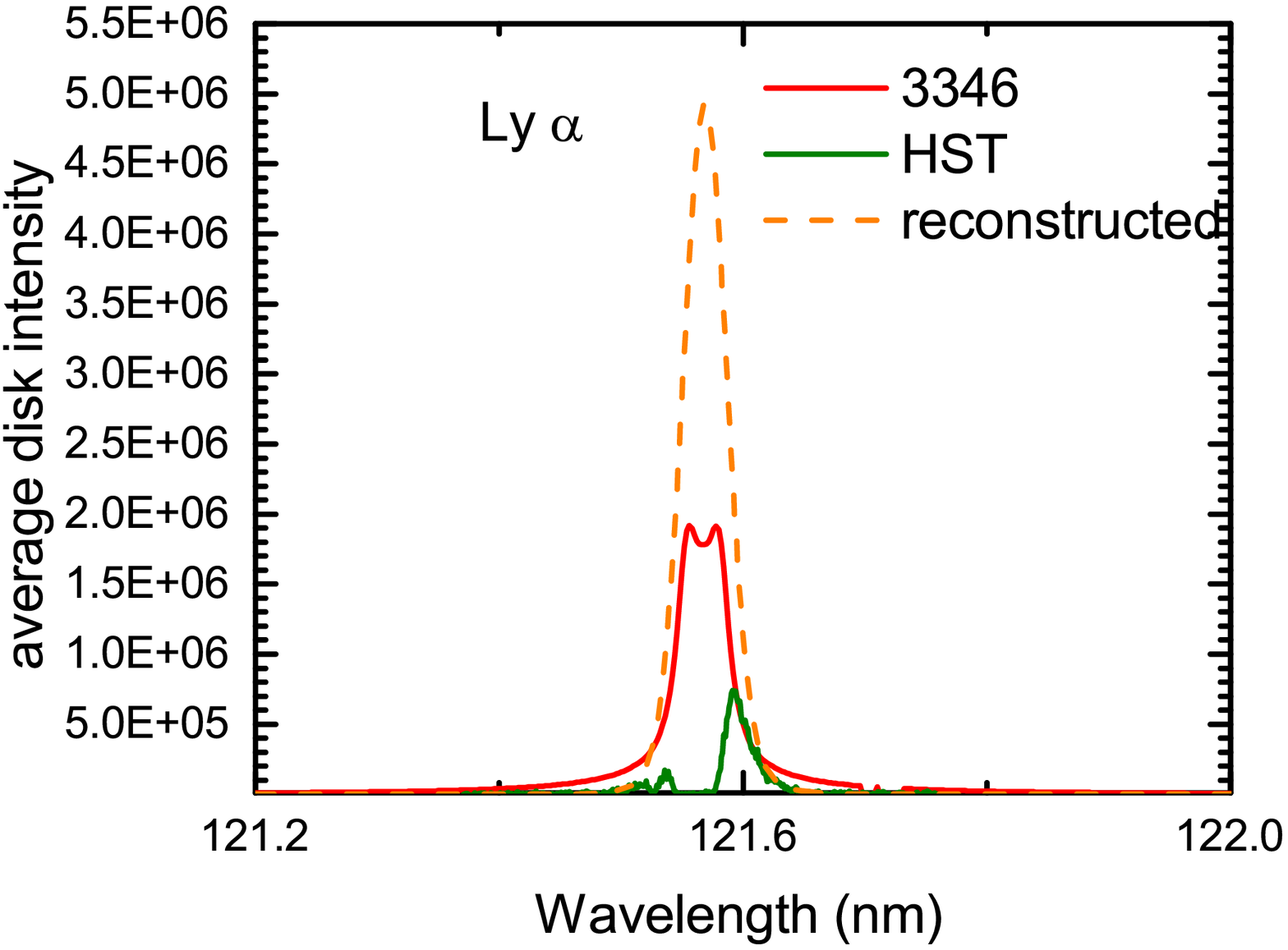}
\caption{Comparison of the GJ~832 Ly-$\alpha$ emission line observed
by {\em HST} (green line) with the model 3346 average disk intensity (red
line) and the observed Ly-$\alpha$ line (orange dashed line) reconstructed
to remove interstellar absorption\label{8}.}
\end{figure}

Table~1 shows the fluxes of observed FUV emission lines at a
distance of 1~AU from the star compared to the spectral synthesis
intensities computed from the  models for GJ~832 and the quiet and active Sun.
For the C~II 133.45 and 133.57~nm lines, we have corrected the observed
fluxes for an assumed 30\% absorption by interstellar C~II,
similar to our correction for the Mg~II lines. The computed fluxes of 
the transition region lines (Si~IV, C~IV, O~IV], and N~V) all lie within 
20\% of the observed fluxes. For the
lines formed in the chromosphere and lower transition region 
(Ly-$\alpha$, Mg~II, Ca~II, and C~II), the agreements with the observed
line fluxes are mostly within 50\%.

The Ly-$\alpha$ line (Figure~8) presents a special case of an 
optically thick
line with broad wings and significant interstellar absorption. We
compare in Figure~8 the model 3346 profile with the observed
profile. Interstellar absorption by H~I Lyman-$\alpha$ and, to a lesser
extent, D~I Ly-$\alpha$ centered at --0.033~nm from the H~I line 
absorb most of
the intrinsic stellar emission line and severely distort its shape. To
obtain a qualitative assessment of the accuracy of the emission line
computed from model 3346, we compare the computed line with the
observed line corrected for interstellar absorption. There are two
techniques now utilized for reconstructing the Ly-$\alpha$ line: (1)
using the interstellar absorption profiles of the D~I Ly-$\alpha$ and
metal lines to infer the H~I absorption \citep{Wood05}, and (2) a
self-consistent solution for the interstellar absorption parameters and the
intrinsic stellar emission line that best fits the observed line
profile \citep{France13,Youngblood16}. Allison Youngblood has kindly
reconstructed the observed Ly-$\alpha$ line (shown in Figure~8) with an 
integrated flux of 3.86 ergs~cm$^{-2}$~s$^{-1}$ at 1 AU using the
technique described in the \citet{Youngblood16} paper. Using an
earlier version of this technique, \citet{France13} obtained an
integrated flux of 5.21 ergs~cm$^{-2}$~s$^{-1}$. By comparison, 
the integrated flux of the computed line is 2.13
ergs~cm$^{-2}$~s$^{-1}$. Given the signal-to-noise of the
observed profile and the uncertainties in the reconstruction technique,
we conclude that the model 3346 Ly-$\alpha$ flux is 
consistent with the reconstructed stellar line flux.

The wings of the computed
Lyman-$\alpha$ line are somewhat broader than the reconstructed
profile. This may result from uncertain atomic data for carbon, because  
C~I bound-free opacity and C~II recombination overlap the Lyman-$\alpha$ 
wings.

There is an interesting trend in the relative line fluxes with
line-formation temperature between GJ~832 and the quiet and active 
Sun. The line 
flux ratios for model 1401 divided by the observed GJ~832
fluxes decrease from 128 for Mg~II, to 12.1 for the combined
C~II, Si~IV, and C~IV lines, to 3.08 for the highest temperature N~V
lines. The corresponding flux ratios for model 1401 divided by model 3346
are similar: 125 for Mg~II, 13.0 for the combined C~II, Si~IV, and
C~IV lines, and 3.0 for the N~V lines. The increased flux of the
high-temperature lines compared to low-temperature lines results from
the steep
temperature rise into the transition region occuring at nearly an order
of magnitude larger pressure in GJ~832 model than in the quiet Sun
model. We note
that a similar displacement of the transition region to higher gas pressures 
occurs in solar models for regions of higher magnetic activity and in
previouly computed active M-dwarf models 
\citep[e.g.,][]{Houdebine97, Walkowicz09}. Models of solar active regions
(e.g., model 1404) have both higher gas pressures, which greatly 
increase the emission line intensities, and steeper temperature gradients 
in the transition region, which partially reduce the increased intensities.

\begin{figure}[h]
\epsscale{0.7}
\plotone{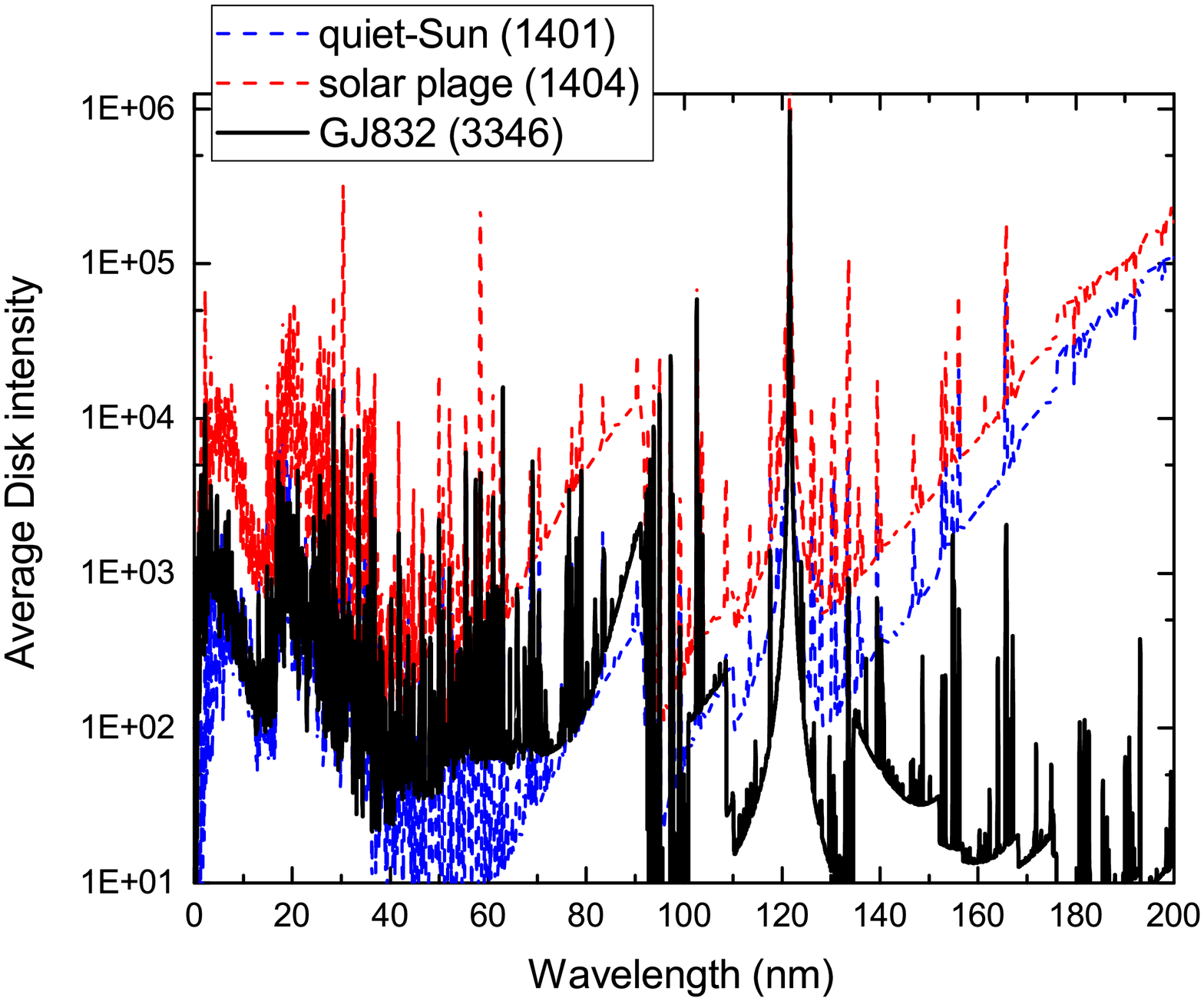}
\plotone{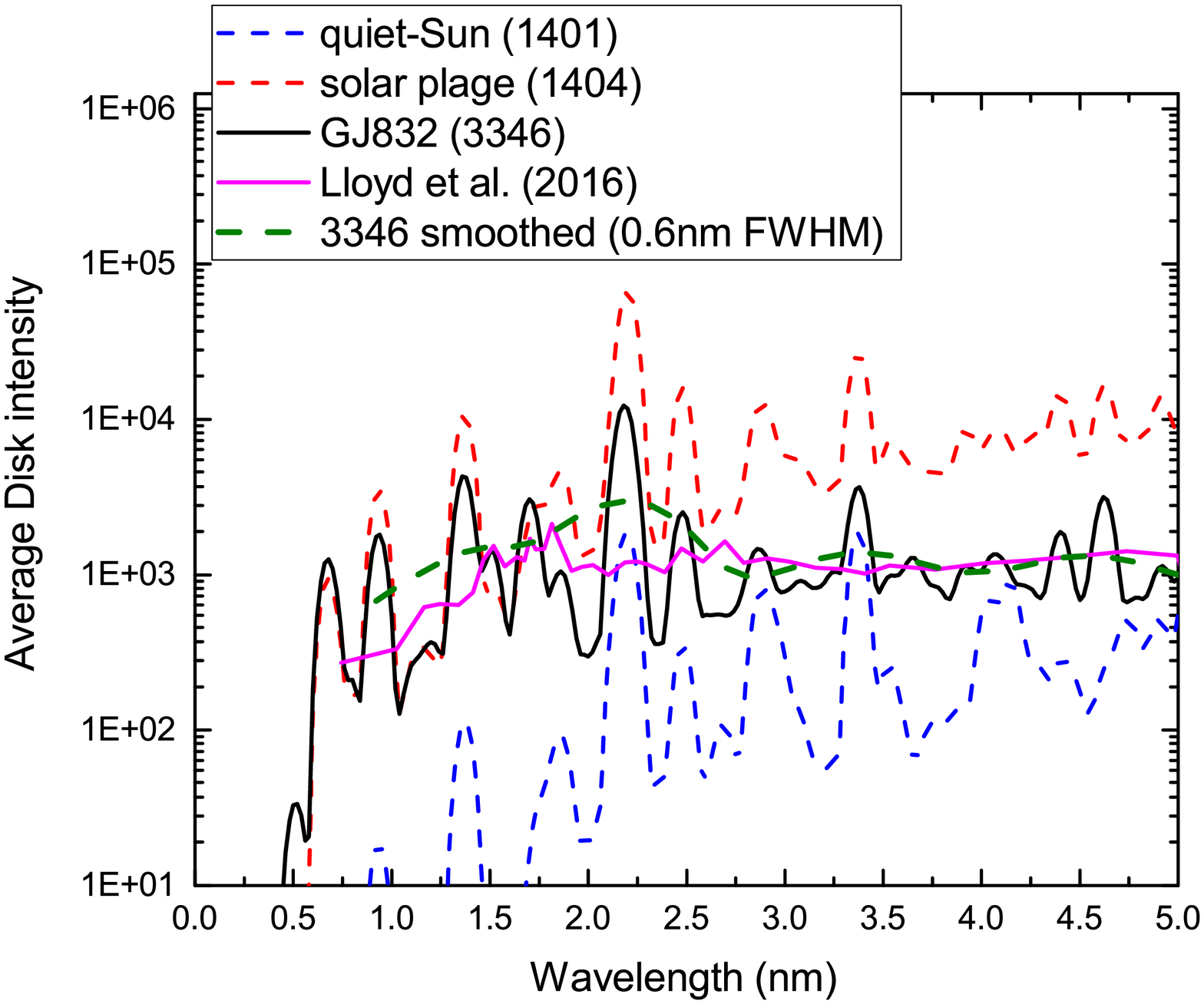}
\caption{{\em Top panel:} The spectral synthesis spectrum of
GJ~832 (model 3346, black line) convolved with a 0.1~nm FWHM cos$^2$
filter. Units of the average disk intensity are 
erg s$^{-1}$ cm$^{-2}$ nm$^{-1}$ sr$^{-1}$.
For comparison, we show spectra of the quiet Sun (model 1401,
blue line) and active Sun (model 1404, red line). 
The brightest emission lines are 
H~I Lyman-$\alpha$ $\lambda$121.567~nm and 
He~II Lyman-$\alpha$ $\lambda$30.391~nm.
{\em Bottom panel:} An expanded plot showing
the observed XMM spectrum of GJ~832 (pink solid line) and model 
spectra for GJ~832 (black line), the quiet Sun (dashed blue line), and 
a solar plage (dashed red line). The dashed green line is the model 
spectrum for GJ~832 smoothed to the approximate resolution of the 
XMM data.\label{fig9}}
\end{figure}

Model 3346 for the transition region of GJ~832 was verified and computed using 
the 0.8 dyne cm$^{-2}$ base pressure and a similar energy deposition and 
flux limiting, as described in Papers~II and III, respectively. 
The resulting model of temperature vs height for the transition 
region lies between solar models 1303 and 1304.

\subsection{The X-ray and EUV Spectra}

Figure~9 shows the complete 0--200~nm spectrum of GJ~832 
computed by adding the 
intensity from the upper layers to that emitted by the lower part 
of the complete model. This procedure assumes that the upper part 
of the model is optically thin at all UV wavelengths, which is a 
fair assumption 
considering the small coronal optical depths. 

The upper transition region and corona of our GJ~832 model
are defined by theoretical considerations of energy balance and
the strong constraints imposed by observations of the C IV (T=100,000~K)
and N V (T=170,000~K)
lines formed in the lower transition region.  
This theoretical method was first described by \citet{Rosner78}
and was addressed by \citet{Fontenla02}
and references therein. \citet{Martens10} then developed
an analytical model consistent with this theoretical method using
parametric forms for the radiative losses and coronal heating. 
In Papers~IV and V we presented a set of models for various regions of the Sun
with models 1401/1311 describing the average
quiet Sun (feature B) and models 1404/1314 describing a moderately active
region of the Sun (feature H). The line intensity ratios in the GJ~832 FUV
spectrum are similar to a slightly less active plage than feature H in
the Sun and are consistent with a pressure of $P=0.8$ dyne cm$^{-2}$ and a
heat flux of $\sim2 \times 10^6$ erg cm$^{-2}$ s$^{-1}$ at 200,000~K.

Our model for GJ~832 is based on the same
theoretical principles as the solar models with the observed
constraints for GJ~832 in a spherically symmetric geometry,
because the corona extends for a significant fraction of the stellar radius.
The resulting detailed structure of the
model is similar to that of the moderate activity
solar plage (feature H in the Sun), except that the model corona 
for GJ~832 is slightly 
hotter and less extended than that of feature H on the Sun. 
These differences may be explained by the higher gravity 
of GJ~832 compared to the Sun. 
We note that \citet{Chadney15} calculated X-ray and EUV
spectra of the
the more active M dwarfs AD~Leo and AU~Mic that are similar to model
spectra of the more active Sun than to the quiet Sun.

The Fe~XII  124.20~nm magnetic dipole line was first detected in the
Sun by \citet{Burton67} and subsequently in $\alpha$~Cen~A (G2~V)
\citep{Pagano04} and $\epsilon$~Eri (K2~V) \citep{Jordan01}. 
This line, shown in Figure~7, forms at a temperature near 1.4~MK (see
the right panel of Figure 11), which is too cool to be in
the corona of our model of GJ832 or in the solar feature
H. Instead, as shown in Fig. 1, this line forms in the upper part of 
the transition region.

\citet{Sanz-Forcada10} reported an X-ray luminosity for GJ~832 of 
$L_X=6.02\times10^{26}$ erg s$^{-1}$ 
integrated over the band 0.1--2.4~keV (i.e., 0.515--12.4~nm) based on
the measured {\em ROSAT} X-ray flux.
This X-ray luminosity corresponds to a surface physical flux of 
$3.88\times10^5$ erg s$^{-1}$ cm$^{-2}$ (assuming a stellar surface area of 
$0.25 A_{\odot}=0.25*1.52\times10^{22} = 3.80\times10^{21}$ cm$^2$) and an 
integrated average disk intensity of $1.55\times10^4$ 
erg s$^{-1}$ cm$^{-2}$ sr$^{-1}$, which is $\sim10$ times that of the quiet Sun 
model 1401 computed in Paper~V. However, the GJ~832 average disk 
intensity in this band is intermediate between 
that of the solar bright network and the weak active-region models 
(1403 and 1404, respectively).

Recently, \citet{Loyd16} observed GJ~832 with the EPIC instrument on 
{\em XMM-Newton} as part of the MUSCLES Treasury Survey Program. They
found that the X-ray spectral energy distribution can be fit by a
two-component model with temperatures $kT_1=0.09^{+0.02}_{-0.09}$ keV
and $kT_2=0.38^{+0.11}_{-0.07}$ keV, corresponding to 
$T_1\approx1.04\times 10^6$ K and $T_2\approx4.41\times 10^6$K. 
The corresponding volume emission measures are $VEM_1\approx2.4\times
10^{49}$~cm$^{-3}$ and $VEM_2\approx0.5\times 10^{49}$~cm$^{-3}$. 
The coronal temperature at the top of our one-component
model lies between these two temperatures and 
$\log L_X=26.26\pm0.05$ ergs~s$^{-1}$\citep{Loyd16}. 
The earlier {\em ROSAT} X-ray luminosity $\log L_X=26.78$ cited by 
\citet{Sanz-Forcada11} is a factor of 3.3
larger than the {\em XMM-Newton} measurement.

The lower panel of Figure~9 compares the observed {\em XMM-Newton} 
spectrum in the 0.7--5.0~nm range 
with model spectra for GJ~832 (model 3346), the quiet Sun (model
1401), and a solar plage (model 1404). Since the resolution of the
EPIC spectrometer on {\em XMM-Newton} is much lower than the
plotted model spectra ($\lambda/\Delta\lambda\approx 5$ in the middle
of the spectral range), the observed spectrum is
smoothed to 0.6~nm FWHM for comparison with the model spectra. The
agreement of the spectral energy distributions of the observed and
model X-ray spectra demonstrates that our model
can be used to approximently represent the X-ray and the unobservable 
EUV spectrum of GJ~832. We note, however, that
coronal emission of more active M dwarfs can be highly variable, and
both the {\em ROSAT} and {\em XMM-Newton} observations were obtained at
different times than the UV measurements.

The integral of the average disk intensity in our model between 
10 and 91.2~nm is $6.21\times 10^{-13}$ erg s$^{-1}$ cm$^{-2}$
sr$^{-1}$, of which 36.4\% is emission from the chromosphere and lower
transition region and 63.6\% is from the upper transition region and
corona. The corresponding flux at Earth is
$f_{\rm  EUV}=6.21\times10^{-13}$  erg s$^{-1}$ cm$^{-2}$ and the
luminosity is $\log L_{\rm EUV}=27.26$ erg s$^{-1}$. From the X-ray
flux observed with {\em ROSAT}, \citet{Sanz-Forcada11} computed
$\log L_{\rm EUV}=27.83$ erg s$^{-1}$. This EUV luminosity is a factor
of 3.7 larger than computed from our model, but the {\em ROSAT} X-ray
flux upon which it is based is a factor of 3.3 larger than the 
{\em XMM-Newton} flux, indicating that GJ~832 was more active at this
earlier time. The scaling between EUV and X-ray luminosity in
eq. 3 of \citet{Sanz-Forcada11} predicts that the EUV luminosity at
the time of the {\em XMM-Newton} observation would be $\log L_{\rm EUV}=27.38$.
A third method for computing the EUV emission is scaling from the 
Ly-$\alpha$ flux \citep{Linsky14}. \cite{Youngblood16} obtained 
$\log L_{\rm EUV}=27.45$, but this
number includes emission in the 91.2--117~nm
region. Excluding the 91.2--117~nm contribution, the fluxes in the
MAST website sum up to $\log L_{\rm  EUV}=27.39$. The close
agreement of the EUV luminosity computed by these three
different techniques (27.26, 27.38, and 27.39) provides confidence that 
all three techniques can provide realistic estimates of the
unobservable EUV luminosity of M dwarf stars.

\begin{figure}[h]
\epsscale{1.1}
\plottwo{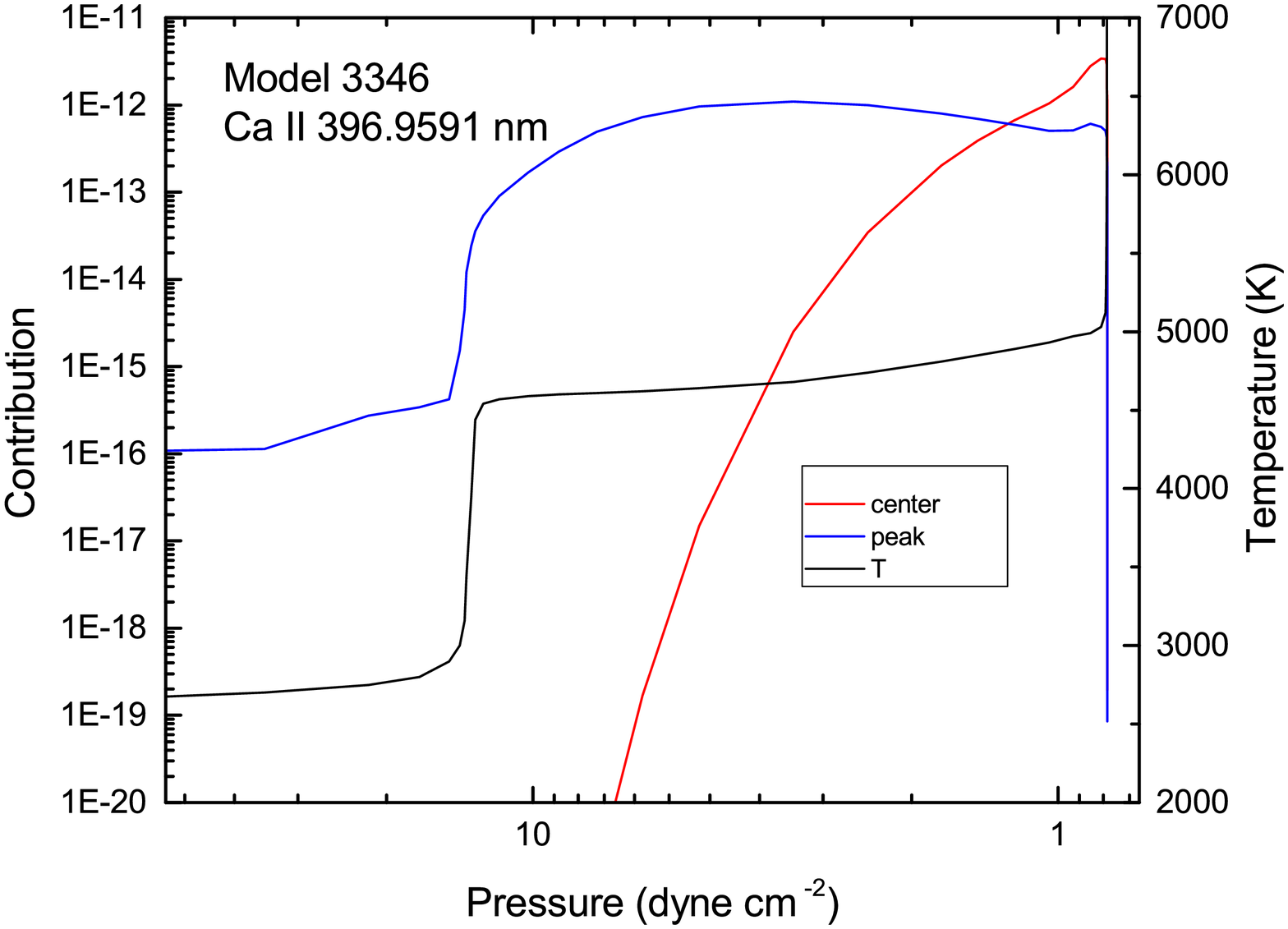}{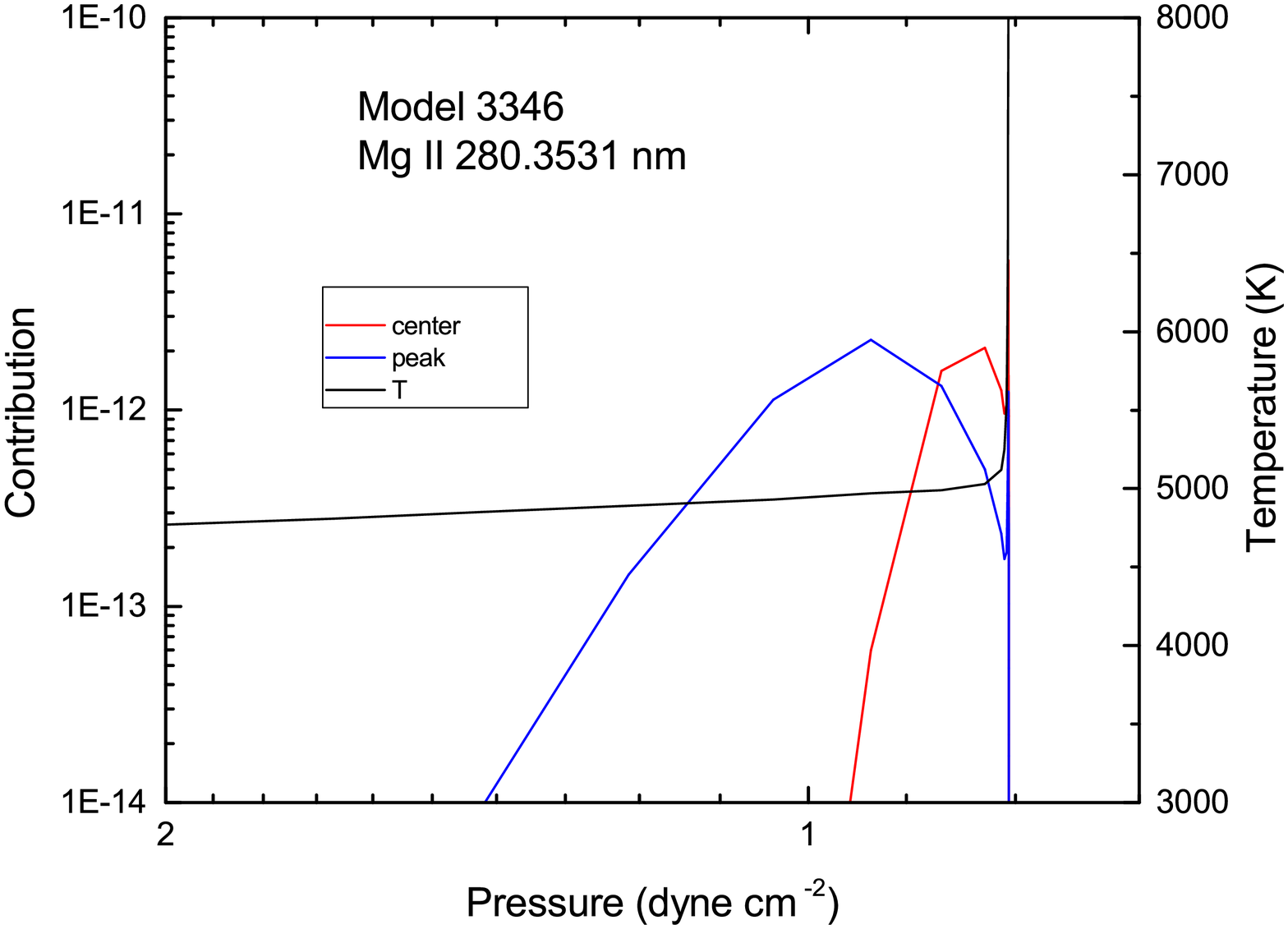}
\caption{ {\em Left panel:} 
Contribution function (attenuated emissivity) at line center (red
line) and emission peaks (blue line) for the Ca~II H and K lines.
 {\em Right panel:} Same for the Mg~II h and k lines.\label{fig10}
Also plotted is the temperature vs. pressure (black line).}
\end{figure}

\section{Spectral distribution of the total stellar flux}

The synthetic spectrum of model 3346 in the top panel of Figure~9 
has an integrated-average disk intensity 
of $1.96\times10^4$ erg s$^{-1}$ cm$^{-2}$ sr$^{-1}$ over the 
0.515--12.4~nm X-ray region, and the model's intensity in the X-ray
band is slightly higher than the 
{\em ROSAT} measurement. The figure also compares the
average disk intensity of model 3346 (GJ~832) to model 1401 (quiet Sun)
and model 1404 (active Sun). We call attention to the increase
in the emission with decreasing wavelength when comparing model 3346 
to the solar models. At 200~nm the average disk intensity for model 3346
is nearly 4 orders of magnitude smaller than for the quiet Sun (model
1401), whereas in the EUV range (30--91.2~nm) the two models have comparable
intensities. At shorter wavelengths ($\lambda<40$~nm)
model 3346 has larger intensities than model 1401. 
Thus the spectral energy
distribution of GJ~832 is ``hotter'' than the quiet Sun and
comparable to the active Sun at $\lambda<91.2$~nm.

Table 2 compares the total flux of model 3346 to the quiet and active
Sun models all evaluated at a distance of 1~AU. The table also compares the
fraction of the model flux in different wavelength bands. The
hotter spectral-energy distribution of GJ~832 is confirmed in this
table by showing that in the 0.2--50~nm band, the fraction of GJ~832's
flux is 30 times larger than the quiet Sun model and also 
larger than the active Sun model. For the FUV band, the 
fraction of the flux for GJ~832 is slightly larger than for the quiet
Sun, but in the NUV band, the fraction of GJ~832's flux is 120 
times smaller.

The middle of the habitable zone (HZ) for a hypothetical exoplanet of
GJ~832 is 0.23~AU, the mean distance between
the ``runaway greenhouse'' and ``maximum greenhouse'' limits described
by \citet{Kopparapu14}. At this distance from its host star, the
exoplanet would see fluxes (0.23)$^{-2}$ = 18.9 times larger than is
shown in Figure~9. As seen from their respective HZs, 
GJ~832 is much fainter than the quiet Sun
in the 150--200~nm band, brighter in the 120--150~nm band, 30 times
brighter than the quiet Sun and comparable to the active Sun (model
1404) in the EUV and 50--120~nm bands, and brighter than
the active Sun model at shorter wavelengths. The UVC band (100--280~nm)
is important for biogenesis processes and DNA damage to life forms on 
planetary surfaces not protected by atmospheric absorption of UV radiation.
Except for major flare events, the smaller radiation over most of the 
UVC band that is received at the surface of a hypothetical exoplanet 
in the HZ of GJ~832 compared to that seen at Earth from the present-day 
Sun could, therefore, place important 
constraints on the origin and evolution of possible life forms
on such an exoplanet \citep[e.g.,][] {Buccino06,Buccino07}.

The NUV flux of GJ~832 at 1~AU is smaller than solar, 
but is substantially larger than what would be expected from an M star 
without a chromosphere. This flux is due to the 
line emissions produced in the upper chromosphere of our model, 
where nonradiative heating takes place. The FUV line ratios 
indicate that the chromosphere-corona transition region of GJ~832 
is more biased to hotter material than that of the quiet Sun. Also, 
the corona of GJ~832 appears to be hotter, but comparable to a 
substantially active Sun. From these trends and our models, we 
find that the EUV flux of GJ~832, which heats the outer
atmospheres of exoplanets and drives their mass loss, is
larger than that of the 
quiet Sun and comparable to a substantially active Sun.

\begin{figure}[h]
\epsscale{1.1}
\plottwo{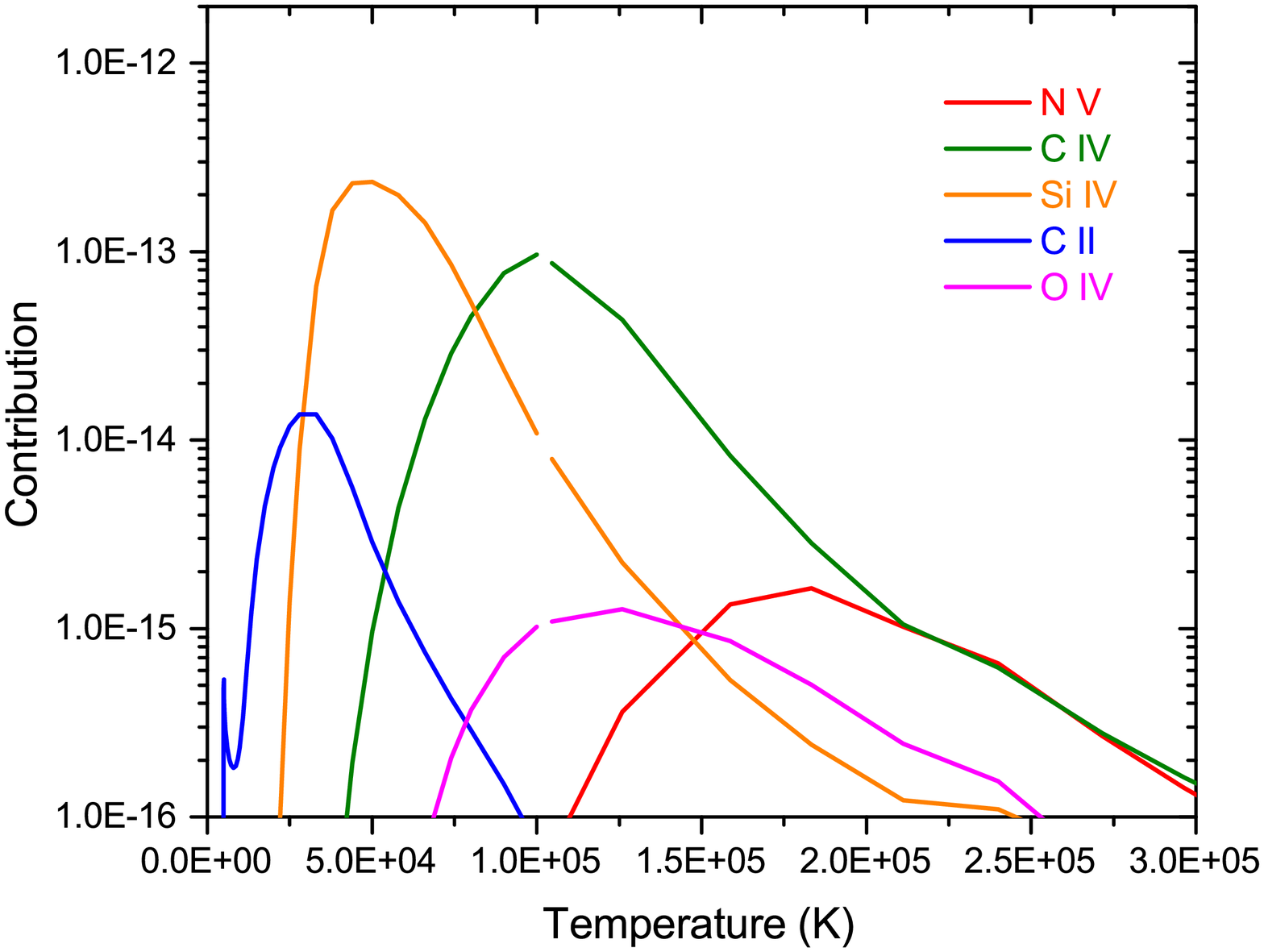}{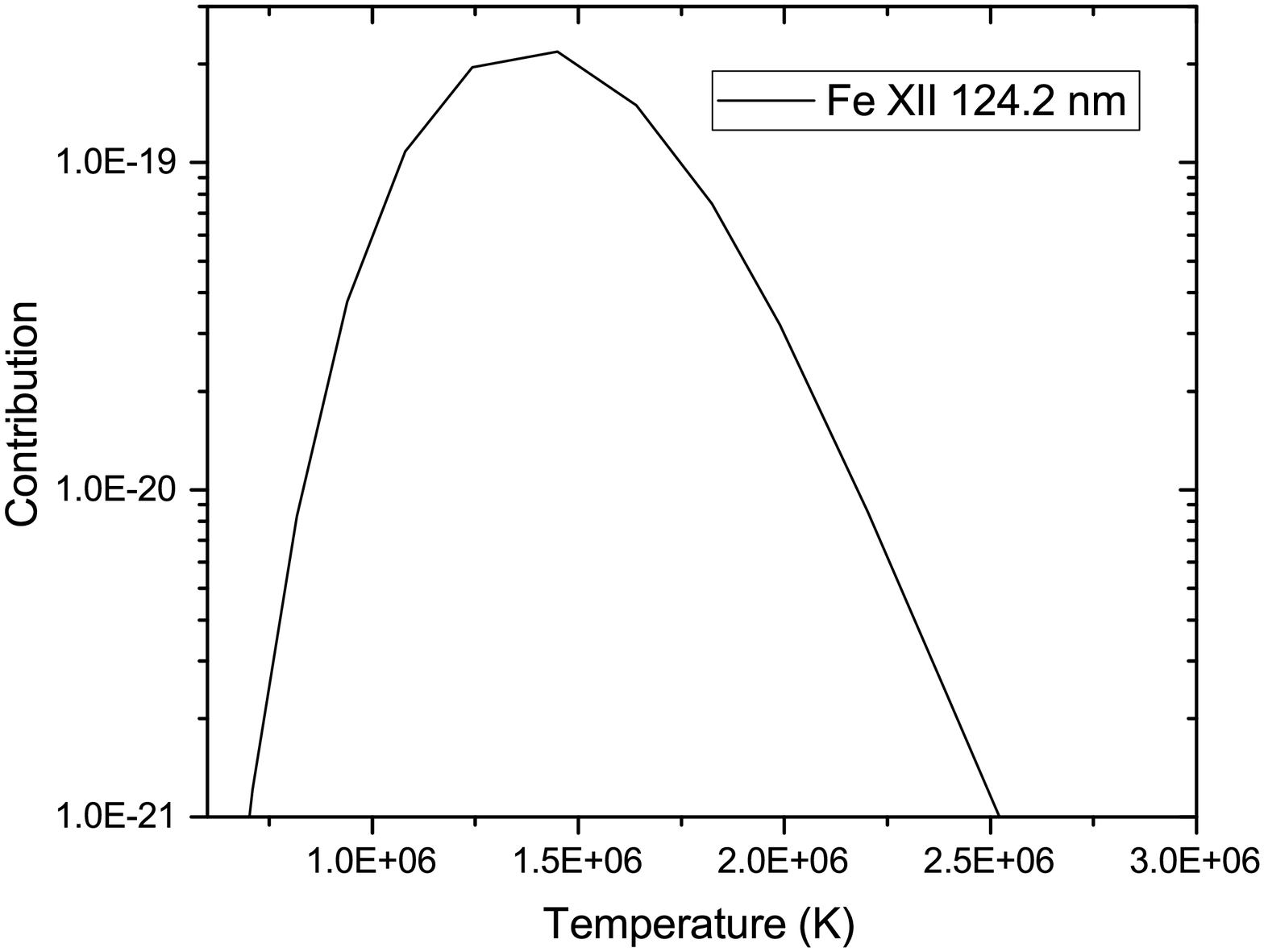}
\caption{{\em Left panel:} Contribution functions for transition region lines. 
{\em Right panel:} Contribution function for the Fe~XII 124.200~nm
  line.}\label{fig11}
\end{figure}

\begin{figure}[h]
\epsscale{1.2}
\plotone{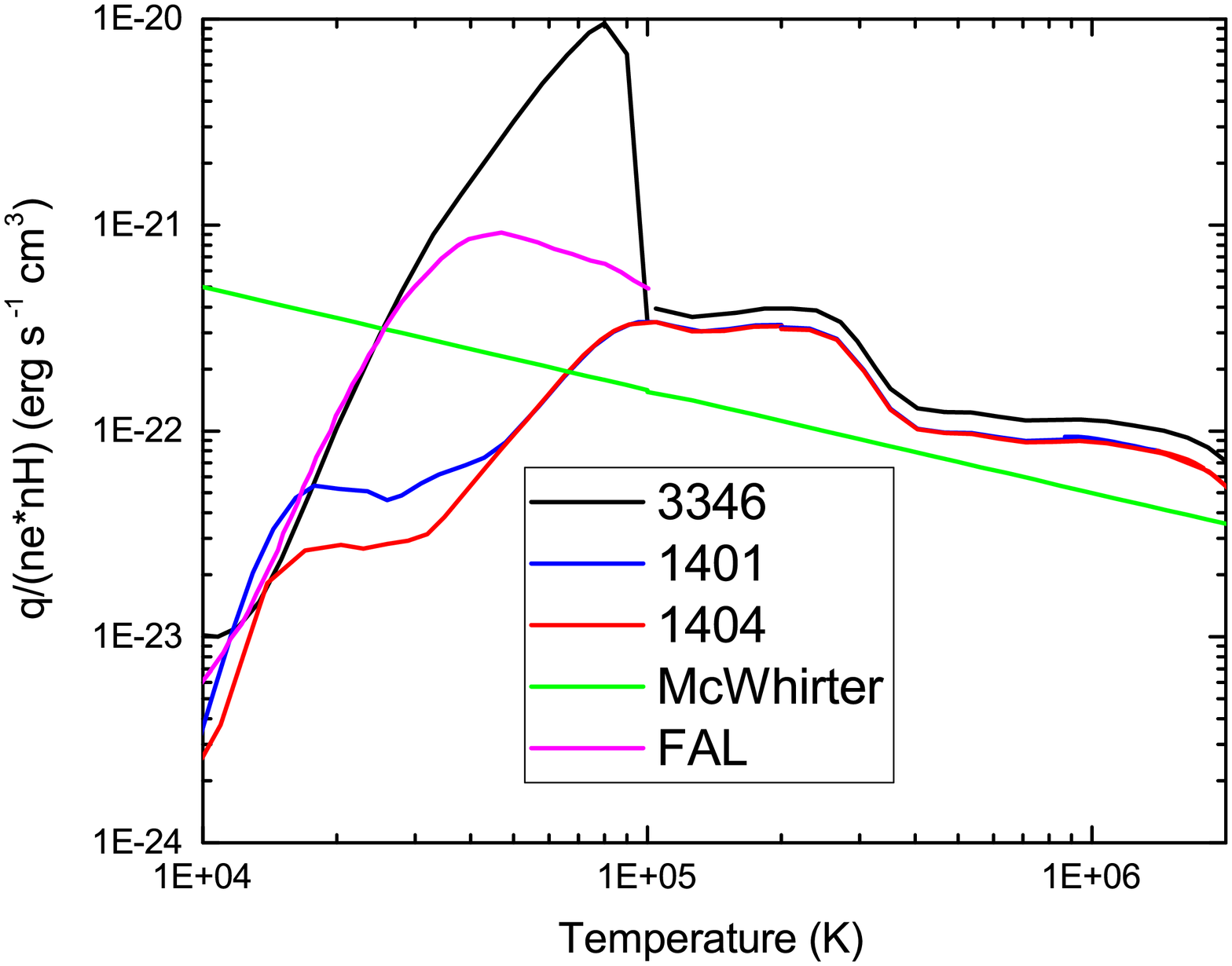}
\caption{Comparison of the net radiative-loss function (NRLF)
  for GJ832 (model 3346, black) with corresponding functions for
  the quiet Sun (1401, blue) and active Sun (1404, red). For
  temperatures above 100,000~K, the NRLFs in the transition
  region and corona are the same for both stars. 
  Below 100,000~K, the NRLFs differ because the steep thermal
  gradient in the transition region of model 3346 induces strong
  diffusion of neutral H to high temperatures where Ly-$\alpha$ is an
  efficient radiator. Also shown is the NRLF 
  (pink) for a solar model computed by \citet{Fontenla02} and a fit
  (green) to the NRLF computed by \citet{McWhirter75}}.\label{fig12}
\end{figure}

\section{DISCUSSION}

\subsection{Contribution Functions}

We calculate contributions functions, i.e., attenuated emissivity, as
defined by Equation~1 in Paper~I.
We show in Figure~10 the contribution functions for the line centers
and peaks of the Ca~II H and Mg~II h lines as a function of pressure
and corresponding temperature in model 3346.
While the centers of these
optically thick lines are formed at relatively high temperatures,
e.g., 10,000 to 20,000~K for Mg~II, the line peaks are formed near
6000--7000~K and the line wings are formed at temperatures as low as 4000~K.

As shown in the left panel of Figure~11, the
centers of the optically thick C~II lines are formed near 25,000~K, but
their line wings are also formed at much cooler temperatures. For the
optically thin Si~IV, C~IV, and N~V lines, there are no wings formed
at cool temperatures. The temperature corresponding to the largest
contribution to the line center emission increases from Mg~I to N~V,
but the contribution functions for all of these species have tails extending
to much higher temperatures, because above the temperature of peak ion
abundance the increase in collisional
excitation rates with temperature partially compensates for the
decrease in ion abundance with temperature.

The broad temperature ranges over which the emission lines are
formed requires that an acceptable atmospheric model must be able to fit many
spectral lines formed over a wide temperature range at the same time. 
Models computed to fit only a limited set of chromospheric diagnostics
(e.g., H$\alpha$ and Ca~II H and K lines) will not be
representative of the entire atmosphere and may not even be an accurate
model for the chromosphere as higher temperature plasma
can contribute to the line center emission.

We used the contributions plots in Figures~10 and 11 to
adjust and fine-tune model 3346 to match the observations.  
For example, reducing the spacing between the temperature-pressure
grid points in the model where an emission line is formed
decreases the total integrated intensity, while 
expanding the distance between points increases 
the intensity.  This adjustment can be done locally with little effect 
on other nearby lines.

The different thermal structures for the GJ~832 and quiet Sun
models are responsible for the different formation conditions of
important ions and spectral lines. For example, unlike the Sun,
the low temperatures in the upper photosphere and lower chromosphere of
GJ~832 result in Ca being almost entirely neutral. As a result, the Ca~II H
and K lines are formed only in a narrow region at higher temperatures 
in the upper chromosphere, and the line wings are narrower than in solar
spectra. The Ca~I absorption lines and TiO bands are much stronger in GJ~832
than in the Sun because of the cool photosphere. The Mg~II h and k
lines are also narrower than in the Sun, because Mg~II is
abundant where the temperature rises steeply in the upper chromosphere and
the line cores are formed, but there is very little Mg~II
in the cool lower chromosphere where the broad wings are formed in
the Sun. The Si~IV, C~IV, and N~V lines are formed at
40,000--170,000~K in the transition region of the GJ~832 model, which
is similar to their regions of formation in solar models.
 
\subsection{Thermal Structure and Emission Line Strength}

\citet{Vernazza81} showed that the strong flux of chromospheric 
and transition region emission lines in the quiet-Sun network
and plage regions can be explained by
semi-empirical models in which the sharp temperature rise in the
transition region occurs deeper in the atmosphere where the pressure
and density are higher. This is confirmed and extended to active
regions as shown by models in \citet{Fontenla02} and Papers I--IV.
If the temperature gradient in the transition
region is the same, the 
effect of higher pressure at a given temperature is to
increase the flux of optically thin lines such as C~IV proportional
$P^2_{\rm TR}$ and to increase the flux of optically thick lines such 
as Ca~II H and K and Mg~II h and k by a smaller amount.
Grids of M-dwarf models built to fit the observed 
Ca~II and H-$\alpha$ lines, including the models of \citet{Hawley92} and 
\citet{Houdebine97},
also show that increasing Ca~II flux and the transition of  H-$\alpha$ 
from an absorption to an emission line naturally follow from thermal
structures with increasing pressure in the upper chromosphere and transition
region.

In our model 3346 for GJ~832, the transition region pressure is about
4 times larger than model 1401 for the quiet Sun (Figure~1), implying
that the transition-region lines on model 3346 should be 16 times 
brighter per unit area on the star. However, Table~1 shows that 
after correction for the 
factor of 4 smaller surface area of GJ~832 compared to the Sun, the
surface fluxes of transition region lines are about the same in both
models, because the transition region of GJ~832 has a much steeper temperature
gradient that compensates for its higher pressure.
The much cooler chromosphere of GJ~832 compared to the quiet
Sun naturally explains why the relative emission of GJ~832 compared to the
quiet Sun increases rapidly from the NUV to shorter wavelengths as
shown in Section~4 and Figure~9. 

In model 3346, the transition region occurs at a pressure $P_{\rm TR}= 0.8$
dynes~cm$^{-2}$, which corresponds to a mass column density, $m=P_{\rm TR}/g$,
$\log m=-4.8$ for the stellar gravity $\log g=4.7$ \citep{Schiavon97}.
This value of $\log m$ is in the middle of the range for the grid of
M1 dwarfs computed by \citet{Houdebine97} and a factor of 2 smaller
than for the quiet M-dwarf model (QT3) computed by \citet{Hawley92}.

\subsection{Net Radiative-Loss Function}

The net radiative-loss rate of a plasma at a given temperature 
(also called the total
radiative-cooling loss) is an essential input
for estimating the energy balance in a wide variety of astrophysical
environments, including quasars, galactic clusters, interstellar medium, and
stellar atmospheres. The ingredients needed for calculating the net
radiative loss as a function of plasma temperature,
which we call the net radiative loss function (NRLF),  include the
assumed abundances, ionization and recombination rates, excitation
rates, the important continuum and line-emission processes, and a
comprehensive catalog of oscillator strengths and 
collisional-excitation rates. 

Early computations of the NRLF
\citep[e.g.,][]{Cox71,McWhirter75,Raymond76} 
included bremsstrahlung, 
recombination radiation, and emission from a limited number of spectral 
lines, but these calculations assumed an optically thin plasma 
in all transitions and no molecules or dust. These functions
showed peak emission at 1--3$\times 10^5$~K with a steep dropoff at
lower temperatures. Subsequent calculations for solar models 
\citep[e.g.,][]{Avrett81} showed the importance of Ca~II and Mg~II in
providing a peak in the NRLF at temperatures between 5,000
and 6,000~K. 
In their non-LTE calculations for a solar model that included 
a large number of spectral lines 
from neutral and singly ionized atoms, \citet{Anderson89} found that the
Fe~II lines contribute about half of the NRLF in the lower
chromosphere. Using the Pandora code, \citet{Vieytes05} and 
\citet{Vieytes09} computed non-LTE models for G and K dwarf stars to fit 
the Ca~II and H$\beta$ lines,
and computed NRLFs in the lower and upper chromospheres.
The NRLF computed in Paper~II 
for the quiet Sun is compatible with those computed by
\citet{Anderson89}, and the NRLFs for active regions on the Sun
are substantially larger. \citet{Avrett15} computed NRLFs for
H~I, Ca~II, Mg~II, and Fe~II in their quiet-Sun and sunspot models, and
\citet{Houdebine10b} computed NRLFs for two M 
dwarfs, GJ~867A and AU~Mic (quiesecent and plage).
 
Figure 12 compares the NRLF for the GJ~832 model with those
for a quiet Sun model (1401) and an active Sun model (1404).
We assume that all spectral lines formed in
the transition region and corona are optically thin, but for all other
lines that may be optically thick we compute in non-LTE the NRLF for 
transitions between two atomic levels, also called the net 
radiative bracket. At temperatures above
100,000~K, the NRLFs of the two stars are the same, because
the emission is from plasmas with the same abundances and all transitions
are optically thin. Below 100,000~K, the functions differ because
fitting the GJ~832 transition region line fluxes requires a much steeper 
thermal gradient in the the transition region than for the Sun,
leading to increased thermal diffusion of neutral H
upward into the transition region and, therefore, stronger 
Ly-$\alpha$ emission \citep[cf.][]{Fontenla02}. 
The thermal gradient in the GJ~832 model is sufficiently steep
that diffusion leads to H being 5\% neutral at 66,000~K.

We also compare in Figure~12 our NRLF with that 
computed by \citet{McWhirter75}, which has often been used for studying the 
energy balence of hot astrophysical plasmas. The often used 
$T^{-1/2}$ fit to the \citet{McWhirter75} NRLF shown 
in Figure~12 lies below the new function at $T>20,000$~K, 
but has a similar slope at temperatures above 100,000~K. 
The new atomic data and many more lines of highly ionized Fe and 
other elements in the 
new models, and the diffusion of neutral H in the transition region,
likely explain why the \citet{McWhirter75} NRLF is low and
should not be used at temperatures 
below $10^5$~K, because it diverges substantially from the
atmospheric models. The figure also shows the NRLF 
computed by \citet{Fontenla02} based on older atomic rates and a high 
abundance of oxygen. The difference with the newer solar and GJ~832 models 
is mainly in the 20,000--100,000~K temperature range where some lines
are optically thick and thermal diffusion of neutral hydrogen is important.

\subsection{Problems fitting the Mg~I 285.2~nm Emission Line}

Our selection of the best temperature-pressure distribution in
the chromosphere was driven by the need to fit the observed Ca~II 
and Mg~II lines. These lines are formed in an overlaping
temperature range with the main contribution for the Ca~II lines 
in the temperature plateau near 4500~K (left panel of Figure~10)
and the Mg~II lines at slightly higher temperatures (right panel 
of Figure~10). There is also a
significant contribution to the formation of the Ca~II lines at
temperatures down to 4000~K where the Mg~I 285.2~nm line is formed. We
found that in order to fit both the Ca~II and Mg~II lines, we needed a
chromospheric thermal structure that produces too much flux in
the Mg~I line. We have explored whether the overly bright Mg~I
emission could be reduced with different chromospheric  
thermal structures, Mg abundance, or
continuum opacity without a solution. We conclude that the bright Mg~I
line likely results from inaccuracies in the atomic ionization and
recombination rates. Our model uses an 
approximate formula for these rates, but \citet{Osorio15} have 
recently computed new rates for
inelastic electron and hydrogen atom collisional ionization of Mg~I 
including charge transfer. Unfortunately, they do not publish rates for
Mg~I ionization or recombination, and in the application of the Mg~I rates 
to a grid of non-LTE model atmospheres \citep{Osorio16} does not include 
these rates.
 
\subsection {The effect of different thermal structures in M dwarf 
and solar models on the formation of the Ly-$\alpha$ line and
electron densities}

The cooler upper chromosphere of GJ~832 does 
not contribute as much to the Ly-$\alpha$ flux
as the warmer upper chromosphere in solar models.
For GJ~832 the line is formed almost
entirely in the transition region through the particle diffusion 
process that allows neutral H to exist at relatively high
temperatures. This process also dominates the emission in 
the solar Ly-$\alpha$ line center, but the peaks and wings of the 
solar line are formed mainly in the upper chromosphere, which is 
not the case for GJ832 because of the cooler temperatures in its 
upper chromosphere.

Another important difference between GJ~832 and the Sun is that the 
electron density in the cooler 
chromosphere of GJ832 is much smaller than in the solar chromosphere. 
In the solar chromosphere, the heavy species are singly ionized 
everywhere, but in GJ~832 they are mostly neutral in the lower
chromosphere and then become singly ionized in the upper chromosphere 
due to the temperature increase and density decrease with height. 
Thus, processes such as the Farley-Buneman instability 
can only occur in the lower chromosphere of GJ~832 at low densities where 
the residual electrons that have a density of $\sim$ 2--3$\times 10^9$ 
cm$^{-3}$ can be heated to collisionally ionize the heavy abundant 
elements (Mg, Si, Fe), increasing 
the electron density to $\sim 10^{10}$ cm$^{-3}$. In the case of the
quiet Sun, the electron density never decreases below $\sim 5\times 10^{10}$
cm$^{-3}$ and rises to about $\sim 10^{11}$ cm$^{-3}$ in the upper 
chromosphere. While the electron densities in GJ~832's chromosphere are about 
ten times smaller than in the Sun, the higher pressure in GJ~832's 
lower transition region results in higher electron densities
comparable to moderately active solar regions.

\section{SUMMARY AND CONCLUSIONS}

This paper describes a complete physical model and synthetic spectrum 
for the M-dwarf star GJ~832 computed to fit the observed UV and optical spectra
and the X-ray flux. This is the first semi-empirical model of an
M star including the photosphere, chromosphere, transition
region, and corona that is computed in full non-LTE using the Solar-Stellar 
Radiation Modeling tools previously employed for computing the most recent 
comprehensive solar-atmosphere models. The GJ~832 model can serve as a
prototype for computing an exoplanet's complete radiation environment,
in particular the unobservable EUV flux,
when observations are not available because of the absence of
suitable telescopes (especially {\em HST}) or our inability to 
observe a host star at earlier times in its evolution.  
We compute very high-resolution synthetic spectra 
($\lambda/\Delta\lambda\sim10^6$)
from X-rays to the infrared for comparison with important diagnostics
including the emission lines of Ca~II, Mg~II, Ly$\alpha$, H$\alpha$
C~II-IV, Si~III-IV, and N~V. We list here our
results and conclusions:

(1) Our model 3346 for GJ~832 differs quantitatively from
the quiet Sun model 1401, but has qualitative
similarities. The photosphere and lower chromosphere of the GJ~832
model are much cooler than the
quiet Sun model at the same pressures with a minimum temperature of
about 2600~K compared to about 3800~K for the quiet Sun. However, the pressure
corresponding to the minimum temperature is not very different for both
models. Above the temperature minimum, both models have a steep
temperature rise to a somewhat curved ``plateau'' near 4700~K 
for GJ~832 and 6300~K for
the quiet Sun. 

(2) The spectral energy distribution of GJ~832 is ``hotter'' than for
the quiet Sun. Comparing the average disk intensity of GJ~832 (model 3346)
to the quiet Sun (model 1401), we find that GJ~832 is nearly $10^4$
times fainter at 200~nm, but comparable to the quiet Sun in the 91--130~nm
spectral region and much brighter than the quiet Sun in the X-ray and EUV
bands (0--91~nm).
The excellent agreement of our computed EUV luminosity with that
obtained by two other techniques indicates that our model predicts 
reliable EUV emission from GJ~832.

(3) Compared to the Sun, GJ~832 may be slightly metal deficient, but
probably not as metal deficient as log[Fe/H]=--0.12 proposed by 
\citet{Johnson09}. The fit to the TiO fluxes
assuming solar abundances is consistent with metal abundances being 
close to solar.

(4) We compute the temperature dependence of the 
net radiative-loss function (NRLF) for model 3346 using 
modern atomic oscillator strengths
and collisional rates, non-LTE radiative transfer, and a 
thermal structure for an M dwarf that
provides the basis for computing future theoretical atmospheric models 
based on energy balance and heating rates.
At temperatures above 100,000~K, the NRLF for GJ~832 is the same
as for quiet and active solar models, because the abundances are the same in
these models and all transitions are optically thin. 
The much steeper temperature gradient in the transition
region of GJ~832 results in strong diffusion of neutral H into the
transition region and a much larger NRLF at
20,000--100,000~K than for the quiet Sun.

While model 3346 does an excellent job of fitting the observed UV and
optical emission lines and continua, our approach to semi-empirical
modelling indicates the need for future work. The computed Mg~I
285.2~nm shows the photospheric absorption more or less correctly but has
a large central emission that is only very weak in the observed profile.
We believe that this discrepancy is the result of incorrect atomic
ionization/recombination rates that allow too much Mg~I 
to be present in the upper chromophere. Although small changes in the 
chromospheric thermal structure between 4000 and 5000~K may also be
significant,
our tests showed that with the current ionization/recombination rates such
changes would lead to too large Mg II emissions before the Mg I emission
becomes small. 
The larger width of the shoulders of the computed Lyman-$\alpha$ emission
line could result from the approximations used in the PRD or from
inaccurate atomic collisional and ionization data for the C~I continua 
that overlap the
Lyman-$\alpha$ wings. We find that the inclusion of molecular
absorption lines (especially TiO but also other diatomics) is
essential for fitting the wings of the Na~I D and H$\alpha$ lines.
Finally, we suggest that the opacity sources in the NUV, and particularly
the H$_2$, need further study.

     While our semi-empirical model for GJ~832 incorporates state-of-the-art
radiative transfer techniques and atomic and molecular rates, it does
not attempt to describe atmospheric inhomogeneity or time variability. 
If the Sun is a useful guide, then M dwarfs should have three-dimensional 
structures, active regions, starspots, flares, and other forms of
spatial and temporal variability. Such phenomena are indeed observed.
Inclusion of these effects will challenge future modelers, but the
present model should be a useful basis for understanding the mean
radiative output of M-dwarf stars that is of great interest for
studying the habitability of exoplanets.
We look forward to a new generation of models for stars
cooler than the Sun that will guide our understanding of stellar
phenomena and physical processes and
the radiation environment of their exoplanets.



\section{ACKNOWLEDGMENTS}

This work is supported by grants HST-GO-13650 and HST-GO-12464 
from the Space Telescope Science Institute to the University of Colorado.
We appreciate the availability of the {\em HST} data through the MAST
Web site hosted by the Space Telescope Science Institute, and stellar
data through the SIMBAD database operated at CDS, Strasbourg, France.
The authors thank the diligent referee for his many helpful suggestions,
Dr. Reiner Hammer, Dr. Svetlana Berdyugina, and Dr. Alexander Brown for 
their helpful comments, and Dr. Allison Youngblood for advice on the 
EUV flux and computing the
reconstructed Ly-$\alpha$ flux. Finally, JLL thanks the Kiepenheuer 
Institut f\"ur Sonnenphysik
for hospitality and an opportunity for shared insights.



{\it Facilities:} \facility{HST (COS)}, \facility{HST (STIS)}, 
\facility{CXO (ASIS)}.

\clearpage

\clearpage

\begin{deluxetable}{lcccc}
\tabletypesize{\small}
\tablecaption{Fluxes\tablenotemark{a} (erg cm$^{-2}$ s$^{-1}$ at 1 AU) 
of observed and computed emission lines}
\tablewidth{0pt}
\tablecolumns{5}
\tablehead{
\colhead{Line and wavelength (nm)} & \colhead{GJ 832} & 
\colhead{GJ 832} & \colhead{quiet Sun} & \colhead{active Sun}\\ 
\colhead{} & \colhead{MUSCLES} & \colhead{Model 3346} & 
\colhead{Model 1401} & \colhead{Model 1405}}
\startdata
Ly-$\alpha$ 121-122 & 3.86\tablenotemark{b}  & 2.13 & 2.44 & 46.1\\
Mg~II k 279.6352 & 0.192\tablenotemark{c} & 0.321 &    30.9 & 114\\
Mg~II h 280.3531 & 0.129\tablenotemark{c} & 0.309 &    29.2 & 106.7\\
Ca~II K 393.4777 & 0.196\tablenotemark{d} & 0.163\tablenotemark{e} 
& -- & --\\
Ca~II H 396.9591 & 0.159\tablenotemark{d} & 0.157\tablenotemark{e} 
& -- & --\\
C~II 133.5708 &  2.47e-3\tablenotemark{c} & 2.60e-3 & 5.52e-2 & 4.01\\
C~II 133.4532 & 1.74e-3\tablenotemark{c} & 1.18e-3 & 5.19e-2 & 4.01\\
Si~IV 139.3757 &       1.46e-3 & 1.33e-3 & 2.81e-3 & 0.266\\
Si~IV 140.2772 &       6.44e-4 & 6.95e-4 & 1.42e-2 & 0.113\\
C~IV 154.8189 &        5.28e-3 & 5.17e-3 & 2.86e-2 & 0.147\\
C~IV 155.0775 &        2.47e-3 & 2.60e-3 & 1.79e-2 & 9.18e-2\\
O~IV 140.1158 &        1.74e-4 & 2.69e-4 & 4.62e-3 & 2.42e-2\\
N~V 123.8823 &         2.26e-3 & 1.74e-3 & 7.14e-3 & 3.89e-2\\
N~V 124.2806 &         1.01e-3 & 8.66e-4 & 2.88e-3 & 2.10e-2\\
Fe~XII 124.20 & $<$1.07e-4 & 7.11e-5 & 5.23e-4 & 1.12e-2\\
\enddata
\tablenotetext{a}{Conversion of intensity to flux at 1~AU assumes
$R=0.499R_{\odot}$ for GJ~832. The angular diameter of GJ~832 at 1~AU is 
$1.691\times10^{-5}$ sr. The angular diameter of the Sun at 1~AU is
$6.7993\times10^{-5}$ sr.}
\tablenotetext{b}{Reconstructed Ly-$\alpha$ line flux from
  A. Youngblood (private communication). The reconstructed Ly-$\alpha$
flux obtained by \citet{France13} from the analysis of the same
observed data is 5.21 erg cm$^{-2}$ s$^{-1}$ at 1 AU.}
\tablenotetext{c}{Intrinsic stellar flux assuming 30\% absorption by 
interstellar Mg~II and C~II.}
\tablenotetext{d}{CASLEO flux measured with a rectangular extraction slit 
0.10097~nm wide for the K line and 0.1274~nm wide for the H line.} 
\tablenotetext{e}{Flux computed from the model 3346 synthetic spectrum using 
a rectangular extraction slit 0.1170~nm wide for the K line and 0.1096~nm 
wide for the H line.} 
\end{deluxetable}

\clearpage

\begin{deluxetable}{lcccccc}
\tabletypesize{\small}
\rotate
\tablecaption{Stellar surface radiative flux\label{tbl-2}}
\tablewidth{0pt}
\tablecolumns{7}
\tablehead{
\colhead{} & \colhead{Total flux at 1AU} & \colhead{Short wave} & 
\colhead{FUV} & \colhead{NUV} & \colhead{Vis/NIR} & \colhead{FIR}\\
\colhead{} & \colhead{0.1 nm -- 100 $\mu$m} & \colhead{0.2--50 nm} &
\colhead{50--200 nm} & \colhead{200--400 nm} & 
\colhead{400 nm -- 1.6 $\mu$m} & \colhead{1.6--100 $\mu$m}
}
\startdata
Units & erg cm$^{-2}$ s$^{-1}$ & relative\tablenotemark{a} & 
relative\tablenotemark{a} & relative\tablenotemark{a} &
relative\tablenotemark{a} & relative\tablenotemark{a}\\
GJ 832\tablenotemark{b} & 4.922e4 & 1.02e-5 & 
  5.85e-5 & 2.26e-3 & 0.642 & 0.356\\
quiet Sun & 1.388e6 & 1.12e-6 & 8.30e-5 & 0.087 & 0.803 & 0.110\\
active Sun & 1.437e6 & 2.46e-5 & 2.67e-4 & 0.098 & 0.794 & 0.108\\ 
\enddata
\tablenotetext{a}{Fraction of total flux in this wavelength interval, 
relative to the flux reported in column 1.}
\tablenotetext{b}{Model 3346.}
\end{deluxetable}


\clearpage





\begin{thebibliography}{}
\bibitem[Auri\`ere(1982)]{Auriere82} Auri\`ere, M.  1982, \aap,
    109, 301

\bibitem[Allard \& Hauschildt(1995)]{Allard95} Allard, F. \&
  Hauschildt, P. H. 1995, \apj, 445

\bibitem[Allard et al.(2010)]{Allard10} Allard, F. et al. 2010, \apj, 540,
  1005

\bibitem[Allard et al.(2001)]{Allard01} Allard F., Hauschildt P. H., 
Alexander D. R., Tamani A., \& Schweitzer A. 2001, \apj, 556, 357

\bibitem[Anderson \& Athay(1989)]{Anderson89} Anderson, L. S. \&
  Athay, R. G. 1989, \apj, 346, 1010

\bibitem[Avrett(1981)]{Avrett81} Avrett, E. H. 1981, in Solar
  Phenomena in Stars and Stellar Systems, ed. R.M. Bonnet \&
  A.K. Dupree, (Dordrecht: D. Reidel Publishing Co.), p. 173
  
\bibitem[Avrett \& Loeser(2008)]{Avrett08} Avrett, E. H. and Loeser,
  R. 2008, \apjs, 175, 229

\bibitem[Avrett et al.(2015)]{Avrett15} Avrett, E., Tian, H., Landi,
  E., Curdt, W., \& W\"ulser 2015, \apj, 811, 87

\bibitem[Baglin(2003)]{Baglin03} Baglin, A. 2003, Ad. Space Res., 31, 345

\bibitem[Bailey et al.(2009)]{Bailey09} Bailey, J., Butler, R. P., 
  Tinney, C. G., et al. 2009, \apj, 690, 743

\bibitem[Batalha et al.(2013)]{Batalha13} Batalha, N. M., Rowe, J. F.,
  Bryson, S. T. et al. 2013, \apjs, 204, 24

\bibitem[Berger et al.(2010)]{Berger10} Berger, E., Basri, G.,
  Giampapa, M. S. et al. 2010, \apj, 709, 332

\bibitem[Bessell(1995)]{Bessell95} Bessell, M. S. \pasp, 107, 672

\bibitem[Blandford et al.(2010)]{Blandford10} Blandford, R. D.,
  Haynes, M. P., Hucra, J. P. 2010, New Worlds, New Horizons in
  Astronomy and Astrophysics, (Washington, D.C.: National Academies Press)

\bibitem[Bonfils et al.(2005)]{Bonfils05} Bonfils, X., Delfosse, X., 
  Udry, S., Santos, N. C., Forveille, T., S\'egransan, D., 2005, \aap, 
  442, 635

\bibitem[Bonfils et al.(2013)]{Bonfils13} Bonfils, X., Delfosse, X., 
  Udry, S., et al. 2013, \aap, 549, A109

\bibitem[Bozhinova, Helling, \& Scholz(2015)]{Bozhinova15} Bozhinova,
  I., Helling, C., \& Scholz, A. 2015, \mnras, 450, 160

\bibitem[Brain et al.(2010)]{Brain10} Brain, D., Barabash, S., 
  Boesswetter, A., et al. 2010, Icarus, 206, 139

\bibitem[Buccino et al.(2006)]{Buccino06} Buccino, A. P., Lemarchand, G. A., 
  \& Mauas, P. J. D. 2006, Icarus, 183, 491

\bibitem[Buccino et al.(2007)]{Buccino07} Buccino, A. P., Lemarchand, G. A., 
  \& Mauas, P. J. D. 2007, Icarus, 192, 582

\bibitem[Buccino et al.(2011)]{Buccino11} Buccino, A. P., D\'iaz, R. F., 
  Luoni, M. L., Abrevaya, X. C., \& Mauas, P. J. D., 2011, \aj, 141, 34

\bibitem[Buccino et al.(2014)]{Buccino14} Buccino, A. P., Petrucci, R., 
   Jofr\'e, E., \& Mauas, P. J. D. 2014, \apjl,781, 9

\bibitem[Burton, Ridgeley, \& Wilson(1967)]{Burton67} Burton,
  W. M., Ridgeley, A., \& Wilson, R. 1967, \mnras, 135, 207

\bibitem[Cincunegui \& Mauas(2004)]{Cincunegui04} Cincunegui, C. \&
  Mauas, P. J. D. 2004, \aap, 414, 699

\bibitem[Chadney et al.(2015)]{Chadney15} Chadney, J. M., Galand,
  M., Unruh, Y. C., Koskinen, T. T., \& Sanz-Forcada, J. 2015,
  Icarus, 250, 357

\bibitem[Cox \& Tucker(1971)]{Cox71} Cox, D. P. \& Tucker, W. H. 1971,
  \apj, 157, 1157

\bibitem[Cram \& Giampapa(1987)]{Cram87} Cram, L. E. \& Giampapa,
  M. S. 1987, \apj, 323, 316 

\bibitem[Des Marais et al.(2002)]{DesMarais02} Des Marais, D. J.,
  Harwit, M. O., Jucks, K. W., Kasting, K. W., Lin, D. N. C., Lunine,
  J. I., Schneider, J., Seager, S., Traub, W. A., \& Wolfe,
  N. J. 2002, AsBio, 2, 153

\bibitem[Doyle et al.(1997)]{Doyle97} Doyle, J. G., Mathioudakis, M.,
  Andretta, V., Short, C. I., \& Jelinsky, P. 1997, \aap, 318, 835

\bibitem[Dressing \& Charbonneau(2015)]{Dressing15} Dressing, C. D.,
  \& Charbonneau, D. 2015, \apj, 807, 45

\bibitem[Engle, Guinan, \& Mizusawa(2009)]{Engle09} Engle, S. G., 
  Guinan, E. F., \& Mizusawa, T. 2009, AIP Conf. Proc., 1135, 221

\bibitem[Evans et al.(1961)]{Evans61} Evans, D. S., Menzies, A., Stoy,
  R. H., \& Wayman, P. A. 1961, Royal Obs. Bull., 48

\bibitem[Fontenla et al.(1999)]{Fontenla99} Fontenla, J., White,
  O. R., Fox, P. A., Avrett, E. H., \& Kurucz, R. L. 1999, \apj, 518,
  480

\bibitem[Fontenla, Avrett, \& Loeser(2002)]{Fontenla02} Fontenla,
  J. M., Avrett, E. H., \& Loeser, R. 2002, \apj, 572, 636

\bibitem[Fontenla et al.(2007)]{Fontenla07} Fontenla, J.,
  Balasubramaniam, K. S., \& Harder, J. 2007, \apj, 667, 1243 (Paper I)

\bibitem[Fontenla et al.(2009)]{Fontenla09} Fontenla, J, M., Curdt,
  W., Haberreiter, M., Harder, J., \& Tian, H. 2009, \apj, 707, 482
  (Paper II)

\bibitem[Fontenla et al.(2011)]{Fontenla11} Fontenla, J. M., Harder,
  J., Livingston, W., Snow, M., \& Woods, T. 2011, \jgr, 116, D20108
  (Paper III)

\bibitem[Fontenla et al.(2014)]{Fontenla14} Fontenla, J. M., Landim
  E., Snow, M., \& Woods, T. 2014, \solphys, 289, 515 (Paper IV)

\bibitem[Fontenla et al.(2015)]{Fontenla15} Fontenla, J. M., Stancil,
  P. C., \& Landi, E. 2015, \apj, 809, 157 (Paper V)

\bibitem[Fleming, Schmitt, \& Giampapa(1995)]{Fleming95} Fleming, T. A., 
  Schmitt, J. H. M. M., \& Giampapa, M. S. 1995, \apj, 450, 401

\bibitem[France et al.(2012)]{France12} France, K., Linsky, J. L.,
  Tian, F., Froning, C. S., \& Roberge, A. 2012, \apjl, 750, L32

\bibitem[France et al.(2013)]{France13} France, K., Froning, C. S.,
  Linsky, J. L., Roberge, A., Stocke, J. T., Tian, F., Bushinsky, R.,
  D\'esert, J.-M., Mauas, P., Vieytes, M., \& Walkowicz, L. M. 2013,
  \apj, 763, 149

\bibitem[France et al.(2016)]{France16} France, K., Loyd,
  R. O. P., Youngblood, A. et al. 2016, \apj, 820, 89

\bibitem[Fuhrmeister, Schmitt, \& Hauschildt(2005)]{Fuhrmeister05} 
  Fuhrmeister, B., Schmitt, J. H. M. M., \& Hauschildt, P. H. 2005,
  \aap, 439, 1137

\bibitem[Gomes da Silva et al.(2012)]{Gomes12} Gomes da Silva, J., 
  Santos, N. C., Bonfils, X., Delfosse, X., Forveille, T., Udry, S., 
  Dumusque, X., \& Lovis, C. 2012, \aap, 541, 9

\bibitem[Gustafsson et al.(2008)]{Gustafsson08} Gustafsson, B.,
  Edvardsson, B., Eriksson, K., Jorgensen, U. G., Nordlund, A. A., \&
  Plez, B. 2008, \aap, 486, 951.

\bibitem[Hauschildt \& Baron(2008)]{Hauschildt08} Hauschildt, P. H. \&
  Baron, E. 2008, \aap, 490, 873

\bibitem[Hawley \& Fisher(1992)]{Hawley92} Hawley, S. L. \& Fisher,
  G. H. 1992, \apjs, 78, 565

\bibitem[Houdebine(2009)]{Houdebine09} Houdebine, E. R. 2009, \mnras,
  397, 2133

\bibitem[Houdebine(2010a)]{Houdebine10a} Houdebine, E. R. 2010a, 
  \aap, 509, A65

\bibitem[Houdebine(2010b)]{Houdebine10b} Houdebine, E. H. 2010b,
  \mnras, 403, 2157

\bibitem[Houdebine(2010c)]{Houdebine10c} Houdebine, E. H. 2010c,
  \aap, 509, 65

\bibitem[Houdebine et al.(1996)]{Houdebine96} Houdebine, E. R.,
  Mathioudakis, M., Doyle, J. G., \& Foing, B. H. 1996, \aap, 305, 209

\bibitem[Houdebine \& Stempels(1997)]{Houdebine97} Houdebine, E. R.,
  \& Stempels, H. C. 1997, \aap, 326, 1143

\bibitem[Johnson(2014)]{Johnson14} Johnson, J. A. 2014, Phys. Today,
  67, issue 10, p. 11

\bibitem[Johnson \& Apps(2009)]{Johnson09} Johnson, J. A. \& Apps, K. 2009, 
  \apj, 699, 933


\bibitem[Jordan et al.(2001)]{Jordan01} Jordan, C., McMurry,
  A. D., Sim, S. A., \& Arulvel, M. 2001, \mnras, 322, L5

\bibitem[Kashyap, Drake, \& Saar(2008)]{Kashyap08} Kashyap, V. L.,
  Drake, J. J., \& Saar, S. H. 2008, \apj, 687, 1339

\bibitem[Kasting, Whitmire, \& Reynolds(1993)]{Kasting93} Kasting, J. F., 
  Whitmire, D. P. \& Reynolds, R. T. 1993, Icarus, 101, 108

\bibitem[Koch et al.(2010)]{Koch10} Koch, D. G., Borucki, W. J., Basri, G., 
  et al. 2010, \apjl, 713, L79

\bibitem[Kopparapu et al.(2013)]{Kopparapu13} Kopparapu, R. K., 
  Ramirez, R., Kasting, J. F. et al. 2013, \apj, 765, 131 

\bibitem[Kopparapu et al.(2014)]{Kopparapu14} Kopparapu, R. K., 
  Ramirez, R. M., SchottelKotte,J., et al. 2014, \apj, 787, L29

\bibitem[Landi et al.(2013)]{Landi13} Landi, E., Young, P. R., 
  Dere, K. P., Del Zanna, G., \& Mason, H. E. 2013, \apj, 763, 86

\bibitem[Lanza(2008)]{Lanza08} Lanza, A. F. 2008, \aap, 486, 1163

\bibitem[Linsky, France, \& Ayres(2013)]{Linsky13} Linsky, J. L., 
  France, K., \& Ayres, T. 2013, \apj, 766, 69

\bibitem[Linsky, Fontenla, \& France(2014)]{Linsky14} Linsky, J. L.,
  Fontenla, J., \& France, K. 2014, \apj, 780, 61

\bibitem[Llama \& Shkolnik(2015)]{Llama15} Llama, J. \& Shkolnik,
  E. L. 2015, \apj, 802, 41

\bibitem[Loyd et al. (2016)]{Loyd16} Loyd, R. O. P.,
  France, K., Youngblood, A. et al. 2016, \apj, 824, 102 

\bibitem[Maldonado et al.(2015)]{Maldonado15} Maldonado, J., Affer,
  L., Micela, G. et al. 2015, \aap, 577, A132
 
\bibitem[Martens(2010)]{Martens10} Martens, P. C. H. 2010, ApJ,
  714, 1290

\bibitem[Mauas(2000)]{Mauas00} Mauas, P. J. D. 2000, \apj, 539, 858 

\bibitem[Mauas et al.(1997)]{Mauas97} Mauas, P. J. D., Falchi, A.,
  Pasquini, L., \& Pallavicini, R. 1997, \aap,  326, 249

\bibitem[McWhirter et al.(1975)]{McWhirter75} McWhirter, R. W. P., 
Thonemann, P. C., \& Wilson, R. 1975, \aap, 40, 63

\bibitem[Miguel et al.(2015)]{Miguel15} Miguel, Y., Kaltenegger, L., 
  Linsky, J. L., \& Rugheimer, S. 2015, \mnras, 446, 345

\bibitem[Mihalas(1978)]{Mihalas78} Mihalas, D. 1978, Stellar
  Atmospheres Second Edition, (San Francisco: W.H. Freeman \& Co.)

\bibitem[Murray-Clay, Chiang, and Murray(2009)]{Murray-Clay09}
  Murray-Clay, R. A., Chiang, E. I. \& Murray, N. 2009, \apj, 693, 23 

\bibitem[Newton et al.(2016a)]{Newton16a} Newton, E. R., Irwin, J.,
  Charbonneau, D., Berta-Thompson, Z. K., \& Dittmann, J. A. 2016, 
  \apj, 821, L19

\bibitem[Newton et al.(2016b)]{Newton16b} Newton, E. R., Irwin, J.,
  Charbonneau, D., Berta-Thompson, Z. K., Dittmann, J. A., \& West,
  A. A. 2016, \apj, 821, 93

\bibitem[Osorio \& Barklem(2016)]{Osorio16} Osorio,Y. \& Barklem,
  P. S. 2016, \aap, 586, A120

\bibitem[Osorio et al.(2015)]{Osorio15} Osorio, Y., Barklem, P. S.,
  Lind, K., Belyaev, A. K., Spielfiedel, A., Guitou, M., \& Feautrier,
  N. 2015, \aap, 579, 530

\bibitem[Pagano et al.(2000)]{Pagano00} Pagano, I., Linsky, J. L., 
  Carkner, L., Robinson, R. D., Woodgate, B., \& Timothy, G. 2000, 
  \apj, 532, 497

\bibitem[Pagano et al.(2004)]{Pagano04} Pagano, I., Linsky,
  J. L., Valenti, J., \& Duncan, D. K. 2004, \aap, 415, 331

\bibitem[P\'erez Mart\'inez, Schr\"oder, and Hauschildt(2014)]{Perez14}
  P\'erez, M. I. P., Schr\"oder, K.-P., \& Hauschildt, P. 2014,
  \mnras, 445, 270

\bibitem[Plez(1998)]{Plez98} Plez, B. 1998, \aap, 337, 495

\bibitem[Poppenhaeger, Robrade, \& Schmitt(2010)]{Poppenhaeger10} 
   Poppenhaeger, K., Robrade, J., \& Schmitt, J. H. M. M. 2010,
   \aap, 515, A98

\bibitem[Poppenhaeger \& Wolk(2014)]{Poppenhaeger14} Poppenhaeger,
  K. \& Wolk, S. J. 2014, \aap, 565, 1

\bibitem[Raymond, Cox \& Smith(1976)]{Raymond76} Raymond, J. C., 
  Cox, D. P., \& Smith, B. W. 1976, \apj, 204, 290

\bibitem[Redfield \& Linsky(2008)]{Redfield08} Redfield, S. \& 
  Linsky, J. L. 2008, \apj, 673, 283

\bibitem[Robinson, Cram, \& Giampapa(1990)]{Robinson90} Robinson,
  R. D., Cram, L. E., \& Giampapa, M. S. 1990, \apjs, 74, 891

\bibitem[Robrade \& Schmitt(2005)]{Robrade05} Robrade, J., \& 
  Schmitt, J. H. M. M. 2005, \aap, 435, 1073

\bibitem[Rosner, Tucker, \& Viana(1978)]{Rosner78} Rosner, R., Tucker, W. H., 
  \& Viana, G. S. 1978, \apj, 220, 643

\bibitem[Sanz-Forcada et al.(2010)]{Sanz-Forcada10} Sanz-Forcada, J.,	
	Ribas, I., Micela, G., Pollock, A. M. T., García-\'Alvarez, D., 
        Solano, E., \& Eiroa, C. 2010, \aap, 511, A8 

\bibitem[Sanz-Forcada et al.(2011)]{Sanz-Forcada11} Sanz-Forcada, J.,	
	Micela, G., Ribas, I., Pollock, A. M. T., Eiroa, C., 
        Velasco, A., Solano, E., \& García-\'Alvarez, D. 2011 
        \aap, 532, A6 

\bibitem[Scalo et al.(2007)]{Scalo07} Scalo, J., Kaltenegger, L., 
  Segura, A., et al. 2007, AsBio, 7, 85 

\bibitem[Schiavon, Barbuy, \& Singh(1997)]{Schiavon97} Schiavon, R. P., 
  Barbuy, B., \& Singh, P. D. 1997, \apj, 484, 499

\bibitem[Schmitt, Fleming, \& Giampapa(1995)]{Schmitt95} Schmitt, J. H. M. M., 
  Fleming, T. A., \& Giampapa, M. S 1995, \apj, 450.392

\bibitem[Seager et al(2009)]{Seager09} Seager, S., Deming, D., \&
  Valenti, J. A. 2009, in Astrophysics in the Next Decade, 
  Astrophysics and Space Science Proceedings (Netherlands: Springer), p. 123

\bibitem[Segura et al.(2010)]{Segura10} Segura, A., Walkowicz, L. M., 
  Meadows, V., Kasting, J., \& Hawley, S. 2010, AsBio, 10, 751 

\bibitem[Shkolnik, Walker, and Bohlender(2003)]{Shkolnik03}
  Shkolnik, E., Walker, G. A. H., \& Bohlender, D. A. 2003, \apj, 597,
  1092

\bibitem[Shkolnik et al.(2005)]{Shkolnik05}
  Shkolnik, E., Walker, G. A. H., Bohlender, D. A., Gu, P.-G., \&
  K\"urster, M. 2005, \apj, 622, 1075

\bibitem[Short \& Doyle(1998)]{Short98} Short, C. I. \& Doyle,
  J. G. 1998, \aap,  336, 613

\bibitem[Stauffer \& Hartmann(1986)]{Stauffer86} Stauffer, J. R., \&
  Hartmann, L. W. 1986, Cool Stars, Stellar Systems and the Sun, in
  Lecture Notes in Physica 254 (Berlin: Springer), 58

\bibitem[Stelzer et al.(2013)]{Stelzer13} Stelzer, B., Marino, A., 
  Micela, G., L\'opez-Santiago, J., \& Liefke, C. 2013, \mnras, 431, 2063

\bibitem[Su\'arez Mascare\~no et al.(2015)]{Suarez15} Su\'arez
  Mascare\~no, A., Rebolo, R., Gonz\'alez Hern\'andez, J. I., \&
  Esposito, M. 2015, \mnras, 452, 2745 

\bibitem[Tarter et al.(2007)]{Tarter07} Tarter, J. C., Backus, P., 
   Mancinelli, R. L. et al. 2007, Astrobiology, 7, 30

\bibitem[Tian et al.(2008)]{Tian08} Tian, F., Kasting, J. F., Liu, H.,
  \& Roble, R. G. 2008, \jgr, 113, 5008
 	
\bibitem[Tian et al.(2014)]{Tian14} Tian, F., France, K., 
  Linsky, J. L., Mauas, P. J. D., \& Vieytes, M. C. 2014, Earth
  Planet. Sci. Lett., 385, 22

\bibitem[van Leeuwen(2007)]{vanLeeuwen07} van Leeuwen, F. 2007, \aap, 474, 653

\bibitem[Vernazza, Avrett, \& Loeser(1981)]{Vernazza81} Vernazza,
  J. E., Avrett, E. H., \& Loeser, R. 1981, \apjs, 45, 635

\bibitem[Vieytes, Mauas, \& Cincunegui(2005)]{Vieytes05} Vieytes, M. C., 
  Mauas, P. J. D., \& Cincunegui, C. 2005, \mnras, 398, 1495

\bibitem[Vieytes et al.(2009)]{Vieytes09} Vieytes, M. C., Mauas, P. J. D., 
  D\'iaz, R. F. 2009, \mnras, 398, 1495

\bibitem[Walkowicz \& Hawley(2009)]{Walkowicz09} Walkowicz, L. M. \&
  Hawley, S. L. 2009, \aj, 137, 3297 

\bibitem[Wedemeyer, Ludwig, \& Steiner(2013)]{Wedemeyer13} Wedemeyer,
  S., Ludwig, H.-G., \& Steiner, O. 2013, AN, 334, 137

\bibitem[West et al.(2015)]{West15} West, A. A., Weisenburger, K. L.,
  Irwin, J., Berta-Thompson, Z. K., Charbonneau, D., Dittmann, J. \&
  Pineda, J. S. 2015, \apj, 812, 3

\bibitem[Wittenmyer et al.(2014)]{Wittenmyer14} Wittenmyer, R. A., 
  Tuomi, M., Butler, R. P. et al. 2014, \apj, 791, 114

\bibitem[Wood et al.(2005)]{Wood05} Wood, B. E., M\"uller, H.-R., Zank, G. P., 
  Linsky, J. L., \& Redfield, S. 2005, \apj, 628, L143

\bibitem[Youngblood et al.(2016)]{Youngblood16} Youngblood, A.,
  France, K., Loyd, R. O. P. 2016, \apj, 824, 101

\end{thebibliography}
\end{document}